\documentclass[prd,
a4paper,
reprint,
amsmath, amssymb,
aps,
superscriptaddress,
nofootinbib,
floatfix
]{revtex4-2}
\usepackage[english]{babel}
\usepackage[T1]{fontenc}
\usepackage[utf8]{inputenc}
\usepackage[normalem]{ulem}

\usepackage{amsthm}
\usepackage{mathtools}
\usepackage{physics}
\usepackage{xcolor}
\usepackage{graphicx}
\usepackage[left=23mm,right=13mm,top=35mm,columnsep=15pt]{geometry} 
\usepackage{adjustbox}
\usepackage{placeins}
\usepackage{csquotes}

\usepackage{bm} 
\usepackage{balance}
\usepackage[pdftex, pdftitle={Article}, pdfauthor={Author}]{hyperref} 
\usepackage{stmaryrd}
\usepackage{tabulary}

\usepackage{txfonts}
\usepackage{orcidlink}

\renewcommand{\leq}{\leqslant}
\renewcommand{\geq}{\geqslant}

\definecolor{Orange}{rgb}{1.0,0.5,0.15}
\definecolor{Blue}{rgb}{0,0.08,0.65}
\definecolor{Red}{rgb}{0.65,0.08,0.05}
\definecolor{Green}{rgb}{0.15,0.45,0.25}
\definecolor{Pink}{rgb}{1.0,0.05,0.5}
\definecolor{bubbles}{rgb}{0.91, 1.0, 1.0}
\definecolor{aquamarine}{rgb}{0.5, 1.0, 0.83}
\definecolor{bubblegum}{rgb}{0.99, 0.76, 0.8}
\definecolor{bluebell}{rgb}{0.74, 0.74, 0.92}
\definecolor{dollarbill}{rgb}{0.72, 0.93, 0.6}

\newcommand{\observed}[1]{\tilde{#1}}
\newcommand{\BNT}[1]{\hat{#1}}
\newcommand{\zb}[1]{z_b^{#1}}

\begin{document}
\title{Cosmic shear nulling as a geometrical cosmological probe: Methodology and sensitivity to cosmological parameters and systematics}

\author{David~Touzeau}
    \email[Correspondence email address: ]{david.touzeau@ipht.fr}
    \affiliation{Universit\'e Paris-Saclay, IPHT, DRF-INP, UMR 3681, CEA, Orme des Merisiers Bat 774, 91191 Gif-sur-Yvette, France}
\author{Francis~Bernardeau}
      \affiliation{Universit\'e Paris-Saclay, IPHT, DRF-INP, UMR 3681, CEA, Orme des Merisiers Bat 774, 91191 Gif-sur-Yvette, France}
      \affiliation{Sorbonne Universit\'e, CNRS, UMR 7095, Institut d'Astrophysique de Paris, 98 bis Boulevard Arago, 75014 Paris, France}
\author{Karim~Benabed}
    \affiliation{Sorbonne Universit\'e, CNRS, UMR 7095, Institut d'Astrophysique de Paris, 98 bis Boulevard Arago, 75014 Paris, France}
\author{Sandrine~Codis}
    \affiliation{Universit\'e Paris-Saclay, Universit\'e Paris-Cit\'e, DAP, UMR 7158, CEA, Orme des Merisiers Bat 709, 91191 Gif-sur-Yvette, France}
    \affiliation{Sorbonne Universit\'e, CNRS, UMR 7095, Institut d'Astrophysique de Paris, 98 bis Boulevard Arago, 75014 Paris, France}


\date{\today} 

\begin{abstract}
Tomographic weak lensing surveys contain intrinsic symmetries that depend solely on the geometric structure of the Universe. These symmetries can be revealed through null tests and verifying their validity provides constraints on cosmological parameters that govern the background evolution—particularly the redshift dependence of the angular diameter distance. This forms the foundation of the tomographic cosmic shear nulling test introduced in this work. We describe how this test can be implemented, what aspects of cosmology it can constrain, and its specific efficiency in doing so. We also assess its sensitivity to astrophysical effects—such as magnification bias and reduced shear corrections—as well as to observational systematics, including errors in the mean redshift of source bins. Our results show that, in a survey configuration comparable to that of Euclid, this null test can yield complementary constraints on key cosmological parameters such as ${\Omega_{\rm m}, {\rm w}_0}$. However, due to its subdominant constraining power compared to standard 3$\times$2pt analysis and through the identification of a required precision of order $10^{-3}$ on the mean redshift of the bins, we conclude that nulling would better be used as a photometric redshift calibration probe or consistency check. In combination with standard weak lensing and galaxy clustering analysis, it would then offer a promising route to better control systematics and improve the precision of future cosmological measurements.
\end{abstract}


\maketitle

\section{Introduction}

Modern observational cosmology and investigations of large-scale structures, dark energy, and dark matter rely on large-scale galaxy surveys such as KiDS \cite{deJong:2012zb,KiDS:2020suj}, DES \cite{DES:2005dhi,DES:2021wwk}, LSST \cite{LSSTDarkEnergyScience:2012kar}, and Euclid \cite{Euclid:2024yrr}, which use cosmic shear as a weak lensing probe. Weak lensing is a large-scale gravitational effect induced by small matter density variations along the trajectory of photons and manifesting through small deformations of galaxy shapes, as opposed to strong lensing which appears through multiple images and Einstein rings. This effect is expressed through a deformation matrix containing a convergence $\kappa$ and a two-component shear $\gamma^\alpha$ which are, by definition, supposed to be small.

The latest generations of surveys pave the way for tomographic explorations, first described in Ref.~\cite{Hu:1999ek}. This strategy allows us to probe different redshifts and therefore different epochs of the expansion of the Universe, thanks to the increased level of precision of these surveys. In particular, the recently launched Euclid \citep{Euclid:2024yrr} satellite opens a new era for weak lensing analysis using tomography thanks to its precision in redshift, the large sky area covered, its shear measurement accuracy, and its high galaxy density. Thus, the latest generation of surveys leads to a new level of precision in the study of notably dark energy, which requires the development of optimal statistical tools and methodologies to extract cosmological information from these rich datasets. In this context, any possible gain in precision, reduction of uncertainties, or further tests of systematics (which can induce biases and inconsistencies in the cosmological inference pipeline) is highly valuable. In this paper, we propose nulling—via the so-called Bernardeau-Nishimichi-Taruya (BNT) transform \cite{Bernardeau:2013rda} applied to cosmic shear maps—as a new cosmological probe, only based on the geometry of space-time.

Standard weak lensing convergence maps get contributions from every gravitational lens all along the line of sight modulated with an efficiency that depends on the background cosmology and peaks at roughly the mean comoving distance. The BNT transform is a simple linear combination of the maps based only on space-time geometry that allows us to build a new tomographic construction for which the transformed convergence maps do not get contributions from low-redshift lenses. Although the BNT transform is not the first nulling method to be developed on the cosmic shear \cite{Huterer:2005un,Joachimi:2010va,DES:2018lpj,DES:2021jzg}, the BNT transform is the only one that puts forward a full nulling property at the level of convergence maps that is fully based on theoretical assumptions.

A now well-explored consequence of this transformation is that it reduces scale mixing in the projection effects. It therefore makes it possible to build observables with reduced sensitivity to small-scale physics including gravitational nonlinearities and poorly controlled baryonic effects. Such a separation of scales is implemented in Refs.~\cite{Taylor:2020zcg,Taylor:2020imc,2025PhRvD.111h3530G} and appeared as an optimal solution to bypass such difficulties. Reference~\cite{Bernardeau:2020jtc} also shows that the BNT transform can give access to baryon acoustic oscillation (BAO) features in cosmic shear observations. This methodology is also implemented for CMB lensing and line intensity mapping in Refs.~\cite{Fronenberg:2023qtw,Maniyar:2021arp}. 

However, in this paper, we propose a completely different approach for the exploitation of the BNT transform. We do not consider it as a way to build better-behaved observables. We consider the nulling property of the BNT transformed maps as a direct manifestation of the geometric structure of space-time, which can therefore be exploited as a fundamental cosmological probe. So, what we propose here is a null test of the BNT transformed cosmic shear maps.

The idea is as follows: BNT transformed cosmic shear maps do not receive contributions from low-redshift lenses and therefore should have vanishing correlation, within noise level, with any low-redshift tracers of the large-scale structure.

The aim of this article is to show that nulling as a new cosmological test is worth implementing in the context of current and upcoming weak lensing surveys with precise enough source redshift distributions.

In Sec.~\ref{sec:observables}, we recap the formalism of the BNT transform and derive the equations governing the proposed null test (including adequate cross-spectra). Section~\ref{sec:specifications} gives the specifications and Euclid-like context of our study. Then, in Sec.~\ref{sec:bispectrum}, we investigate the effect of higher-order corrections such as the reduced shear and magnification bias, by means of bispectrum corrections. Section~\ref{sec:constraints} gives the resulting cosmological constraints, correlations, and degeneracies of the system. Finally, in Sec.~\ref{sec:photometric}, we investigate photometric redshift errors as a systematic effect.

In Sec.~\ref{sec:specifications}, we also address the similarities and differences of our approach with the so called shear ratio approach, introduced in Ref.~\cite{Jain:2003tba} and applied in Refs.~\cite{DES:2018lpj,DES:2021jzg}, which exploits some partial nulling properties.

\section{Nulling cosmic shear observables}\label{sec:observables}

\subsection{Nulling tomographic cosmic shear maps}

The BNT transform was introduced in Ref.~\cite{Bernardeau:2013rda} as a nulling technique to build linear combinations of cosmic shear maps that are only sensitive to a restricted redshift range. This procedure is purely geometrical and thus scale independent and independent of the power spectrum or matter density contrast. It can be thought of as a way to reorganize the information contained in cosmic shear observables so as to separate information between the maps, hence disentangling the small physical scales at low redshifts from the large physical scales at larger redshifts within a given light-cone aperture.

In a standard tomographic analysis of weak lensing, we work with a set of convergence maps in redshift bins with indices $i=1,\dots,N_z$. The binned convergence, to first order in the matter density contrast $\delta_{\rm m}$ (higher orders are discussed in Sec.~\ref{ssec:corr}), is given by
\begin{equation}\label{eq:convergence}
    \kappa_i (\mathbf{n}) = \int_0^\infty \dd\chi\, w_i(\chi) \delta_{\rm m}(\chi,\mathbf{n}),
\end{equation}
where $\mathbf{n}$ is the position of the source galaxy on the sky and the lensing efficiency in bin $i$ reads $w_i(\chi)=\int \dd z_{\rm S}\, n_i(z_{\rm S}) w(\chi(z_{\rm S}),\chi)$. Here, $n_i(z_{\rm S})$ is the normalized redshift distribution of galaxies in the $i$th tomographic bin as described in Sec.~\ref{sssec:sources} below. The usual lensing kernel is defined as
\begin{equation}\label{eq:kernel}
    w(\chi_{\rm S},\chi)=\frac{3 \Omega_{\rm m} H_0^2}{2 c^2} \frac{D_K(\chi_{\rm S}-\chi)D_K(\chi)}{D_K(\chi_{\rm S}) a(\chi)} \Theta(\chi_{\rm S}-\chi),
\end{equation}
where $H_0$ is the Hubble constant, $c$ is the speed of light, $\Omega_{\rm m}$ is the matter content of the Universe at redshift $0$ and $D_K$ is the comoving distance. It reads
\begin{equation}
D_K(\chi)=
	\begin{cases}
	\frac{c}{H_0} \frac{\sin(\sqrt{K}\chi H_0/c)}{\sqrt{K}}  &$for $ K>0, \\
	\chi &$for $ K=0, \\
	\frac{c}{H_0} \frac{\sinh(\sqrt{-K}\chi H_0/c)}{\sqrt{-K}} \ \ \ &$for $ K<0,
	\end{cases}
 \end{equation}
where $K$ is the curvature and $\chi$ is the usual flat comoving distance
\begin{equation}\label{eq:chiz}
    \chi(z)=\int_0^z \frac{c \dd z'}{H(z')}.
\end{equation}

Note that in Eq.~\eqref{eq:kernel} the redshift of the source $z_{\rm S}$ [respectively $\chi_{\rm S}=\chi(z_{\rm S})$] must be higher than the redshift of the lens $z$ [respectively $\chi=\chi(z)$], hence the Heaviside step function.

Within the BNT transform formalism, from the $N_z$ maps, we build linear combinations to find the effective lensing kernels,\footnote{We use Einstein notation: implicit summation on repeated indices.}
\begin{equation}
    \BNT{w}_a=\sum_i p_{ai}w_i=p_{ai}w_i,
\end{equation}
such that $\BNT{w}_a$ vanishes everywhere except in bins $a-2$, $a-1$, and $a$, for all $a \geq 3$. For the first two maps, we define $\BNT{w}_1=w_1$ and $\BNT{w}_2=w_2-w_1$. We also choose to normalize the coefficients such that $p_{ii}=1$. We note that the transformation is the same for the convergence $\kappa$ and the shear $\gamma$,
\begin{equation}
    \BNT{\kappa}_a= p_{ai}\kappa_i \ , \ \ \ \hat{\gamma}_a^\alpha= p_{ai}\gamma_i^\alpha,
\end{equation}
where $\alpha=1$ or $2$. These nulling coefficients depend on the two quantities
\begin{equation}\label{eq:moments}
n_i^{(0)}=\int \dd\chi \, n_i(\chi) \quad {\rm and} \quad n_i^{(1)}=\int \dd\chi\, \frac{n_i(\chi)}{F_K(\chi)}
\end{equation}
through the conditions
\begin{equation}
\label{eq:constr}
\sum_{i=a-2}^a p_{ai} n_i^{(0)}=0 \quad {\rm and} \quad \sum_{i=a-2}^a p_{ai} n_i^{(1)}=0,
\end{equation}
where we define
\begin{equation}
F_K(\chi)=
	\begin{cases}
	  \frac{\tan(\sqrt{K}\chi H_0/c)}{\sqrt{K}}  &$for $ K>0, \\
	\chi \frac{H_0}{c} &$for $ K=0, \\
	\frac{\tanh(\sqrt{-K}\chi H_0/c)}{\sqrt{-K}} \ \ \ &$for $ K<0.
	\end{cases}
 \end{equation}
Note that in the flat case both $D_K$ and $F_K$ are the standard angular distance (up to the constant factor $H_0/c$), while they differ otherwise. One can see Ref.~\cite{Bernardeau:2013rda} for details of $p_{ai}$ derivations in general.

\begin{figure}[!ht]
   \begin{center}
    \includegraphics[width=\columnwidth]{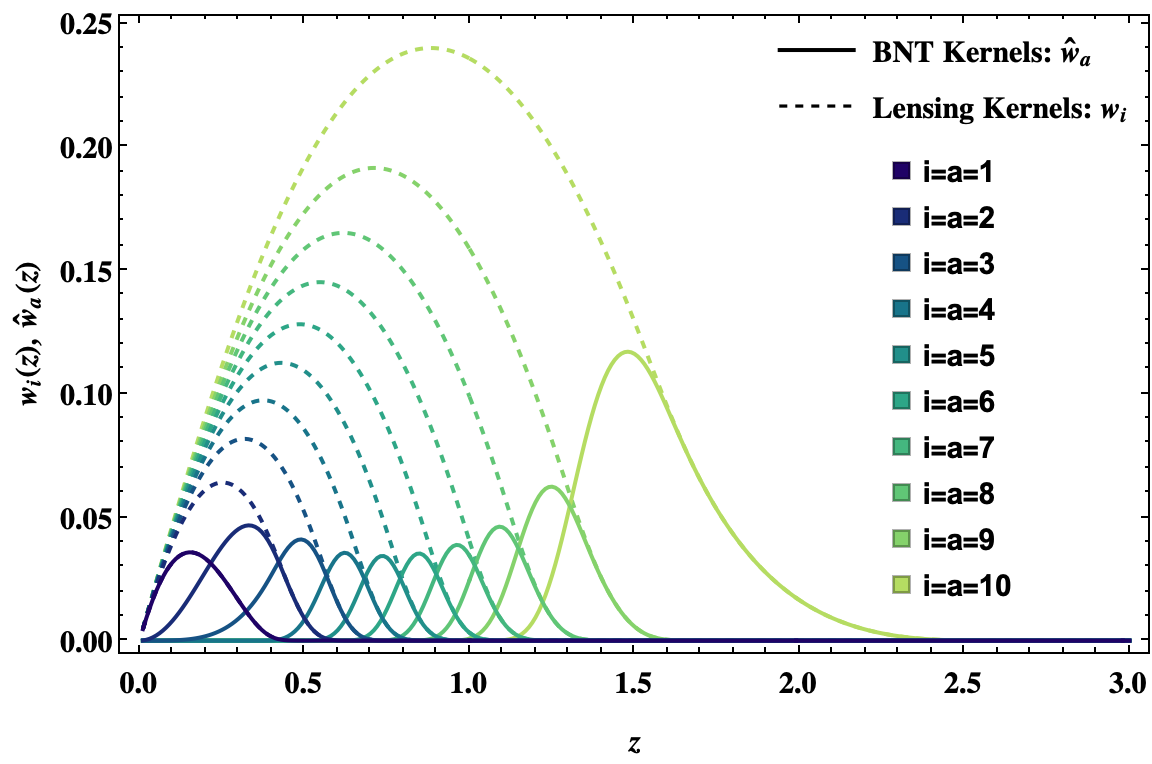}
  \end{center}
  \caption{Shapes of the nulled lensing kernels $\BNT{w}_a(z)$ (solid lines) and of the original lensing kernels $w_i(z)$ (dashed lines), as constructed for 10 tomographic bins and a photometric redshift uncertainty $\sigma_z=0.02$. One should note that $\BNT{w}_1=w_1$.}
  \label{fig:BinsNulling}
\end{figure}

As an illustration, Fig.~\ref{fig:BinsNulling} shows the lensing kernels before and after BNT transformation for 10 equally populated redshift bins described in detail in Sec.~\ref{sssec:sources} below. Lensing kernels after BNT transformation (solid lines) have a compact support compared to the original kernels (dashed lines) that extend all the way from the observer to the source. Each nulled kernel $a$ typically overlaps with only two neighbors on each side from $a-2$ to $a+2$. This illustrates the nulling property of the BNT transform. Note that in this example the nulling coefficients are given by
{\footnotesize \begin{equation} \nonumber 
p_{ai}=
\begin{pmatrix}
	1 & 0 & 0 & 0 & 0 & 0 & 0 & 0 & 0 & \!0 \\
	-1 & 1 & 0 & 0 & 0 & 0 & 0 & 0 & 0 & \!0 \\
	0.24 & \!\!\! -1.24\! & 1 & 0 & 0 & 0 & 0 & 0 & 0 & \!0 \\
	0 & 0.59 & \!\!\!-1.59\! & 1 & 0 & 0 & 0 & 0 & 0 & \!0 \\
	0 & 0 & 0.71 & \!\!\!-1.71\! & 1 & 0 & 0 & 0 & 0 & \!0 \\
	0 & 0 & 0 & 0.79 & \!\!\!-1.79\! & 1 & 0 & 0 & 0 & \!0 \\
	0 & 0 & 0 & 0 & 0.85 & \!\!\!-1.85\! & 1 & 0 & 0 & \!0 \\
	0 & 0 & 0 & 0 & 0 & 0.92 & \!\!\!-1.92\! & 1 & 0 & \!0 \\
	0 & 0 & 0 & 0 & 0 & 0 & 1.01 & \!\!\!-2.01\! & 1 & \!0 \\
	0 & 0 & 0 & 0 & 0 & 0 & 0 & 1.29 & \!\!\!-2.29\! & \!1
	\end{pmatrix}
\end{equation}}
and one can clearly see that the two first bins are specific with $\BNT{w}_1=w_1$ and $\BNT{w}_2=w_2-w_1$.

\subsection{Nulling as a probe of the geometry}\label{ssec:nulling}

In this section, we aim at investigating how the BNT transform coefficients are sensitive to cosmology. In this discussion, we consider normalized equally populated bins such that
\begin{equation}
    n_i^{(0)}=\int \dd\chi\, n_i(\chi)=1.
\end{equation}
From there, solving the constraints given by Eq.~\eqref{eq:constr} yields
\begin{align}
p_{i,i-2}&=\frac{n^{(1)}_{i-1}-n^{(1)}_i}{n^{(1)}_{i-2}-n^{(1)}_{i-1}}, \\
p_{i,i-1}&=\frac{n^{(1)}_i-n^{(1)}_{i-2}}{n^{(1)}_{i-2}-n^{(1)}_{i-1}}=-1-p_{i,i-2}.
\end{align}

Here, one can see that the only dependence of these coefficients on cosmology enters through the redshift dependence of $H(z)$, which is required to compute $\chi$ and $F_K(\chi)$. Those coefficients are, however, independent of $H_0$, which enters as a common prefactor and also completely independent of the matter power spectrum.\footnote{A modeling of $P(k)$ will be introduced in Sec.~\ref{ssec:crossspectra} to illustrate the performance of the proposed scheme, but the test, and its actual performance, is fully independent of any such modeling.} In other words, the BNT transformation matrix is independent of the scale and of the regime of gravitational instabilities. This is the key that allows us to exploit weak lensing data at small scales and in regimes where predictions and numerical experiments are approximate due to poor resolution or unknown small-scale effects. And the fact that it is totally independent of both $H_0$ and $\sigma_8$, two cosmological parameters that are subject to tensions between different measures \citep{Abdalla:2022yfr}, should be viewed as an advantage as it makes it possible to separate pure geometrical effects from the modeling of large-scale structure. 

Let us now introduce a note of caution.
Although this nulling test probes the relation $\chi(z)$, nulling alone cannot fully determine this relation—even if the coefficients $p_{ai}$ were perfectly known over a wide redshift range. More precisely, it leaves degeneracies between cosmological models that pass the test. In Appendix~\ref{sec:proofdegen}, we detail how to identify these degeneracies and demonstrate that they are the only physically relevant ones.

The idea we develop is that if the nulling procedure were implemented using arbitrarily small redshift bins, the nulling coefficients, to first order in the bin width $\Delta z$, would take the form
\begin{equation}
    p_{i-2,i}=1+\frac{\dd^2\xi_K/\dd z^2}{\dd\xi_K/\dd z}\Delta z-\frac{1}{n(z)} \frac{\dd n(z)}{\dd z}\Delta z,
\end{equation}
where $\xi_K=1/F_K$. There are two sources of degeneracies here. The first one comes from the redshift distribution of galaxies, which is to be seen as an observational source of degeneracy. As it is measured with good precision from the survey data and is not the primary focus of our probe, it will not be explored further for now. One should, however, note that any observational effect or systematics of the experiment, such as the magnification limit, will only affect the BNT transform through the source galaxy distribution $n(z)$. The other degeneracy comes from the dependence on the background cosmology with the term ($\xi_K''/\xi_K'$). This means that two cosmological models with the same ($\xi_K''/\xi_K'$) would give the same BNT transform and could not be distinguished by the use of nulling. In other words, nulling coefficients are a probe of the geometry of space-time that cannot help us distinguish between two models which are related by
\begin{equation}\label{eq:degeneracy}
   \left(\frac{\xi_K''}{\xi_K'}\right)^{\rm model1}=\left(\frac{\xi_K''}{\xi_K'}\right)^{\rm model2}.
\end{equation}

\begin{figure}[!ht]
   \begin{center}
    \includegraphics[width=\columnwidth]{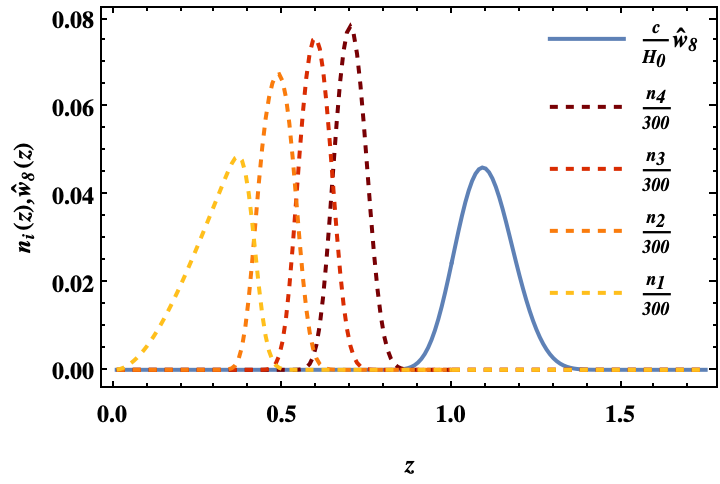}
  \end{center}
  \caption{Eighth nulled lensing kernel (blue solid line) and low-redshift galaxy distributions for bins $1$ to $4$ (dashed lines as labeled).}
  \label{fig:datavector}
\end{figure}

\subsection{Galaxy-nulled shear cross spectra}\label{ssec:crossspectra}

Nulling is a property of BNT transformed maps. Testing whether it is satisfied in a given dataset then requires the definition of a specific estimator. In practice, nulling will be verified if correlations between convergence $\BNT{\kappa}_a$ and any low-redshift tracer vanish.\footnote{We choose here to restrict ourselves to two-point correlators It may not be the optimal choice in terms of constraining power, but is straightforward to implement since it is based on data that are already used for weak lensing analysis. Moreover, it allows us to evaluate the signal to noise formally without the use of simulations.} As we already work with galaxy distributions, a natural choice for these tracers is the distribution of low-redshift galaxies. In the case of probe combinations when working with the 3$\times$2pt method \cite{EUCLID:2020jwq}, it reduces nulling to an almost costless added product since the galaxy-shear cross-spectra are already used and the BNT transform is linear.  As an illustration, we display in Fig.~\ref{fig:datavector} the eighth nulled lensing kernel and the galaxy distributions in the first four redshift bins.

Hence, the observables we focus on in this paper are the cross-spectra, $\mathcal{C}^{\BNT{\kappa} {\rm g}}_{ai}$, defined as
\begin{equation}\label{eq:crossspectra}
    \mathcal{C}^{\BNT{\kappa} {\rm g}}_{ai}(\ell)\delta_D(\boldsymbol{\ell}+\boldsymbol{\ell '})=\left<\BNT{\kappa}_a(\boldsymbol{\ell})\delta^{\rm g}_i(\boldsymbol{\ell '})\right>,
\end{equation}
between nulled lensing maps, $\BNT{\kappa}_a$, and binned low-redshift galaxy distributions, $\delta^{\rm g}_i$, which will be described in detail in Sec.~\ref{sssec:lowzg}. Under the Limber approximation, we can compute the spectrum as
\begin{equation}\label{eq:limbercrossspectra}
    \mathcal{C}^{\BNT{\kappa} {\rm g}}_{ai}(\ell)=\int \frac{\dd \chi}{\chi^2} \BNT{w}_a(\chi) n^{\rm g}_i(\chi) P\left(\frac{\ell}{\chi},z(\chi)\right),
\end{equation}
where $P$ is the matter power spectrum. We focus on the case $i\leq a-3$ such that the respective kernels $\BNT{w}_a$ and $n^{\rm g}_i$ are expected not to overlap, and hence $\mathcal{C}^{\BNT{\kappa} {\rm g}}_{ai}(\ell)$ to vanish.

Note that intrinsic alignments (IA) are expected to preserve the nulling property. At this level of description, in standard IA modeling \cite{Joachimi:2009ez}, the IA kernel is restricted to the location of the sources since it is mainly a local effect due to tidal effects. This remains the case after the BNT transformation, which only mixes adjacent bins as illustrated by the $p_{ai}$ displayed in Eq.~\eqref{eq:constr}, thereby ensuring the preservation of the nulling property.\footnote{From a mathematical point of view, this intrinsic alignment contribution would come as an additional term in Eq. 6 namely $\hat{\gamma}_a^\alpha= p_{ai}\gamma_i^\alpha+p_{ai}e_i^\alpha$ with $e_i^\alpha=\int\dd z_{\rm S}\, n_i^g(z_{\rm S}){\cal A}(z_s)$ the intrinsic alignment contribution that only depends on random fields at the location of the sources. Since $p_{ai}$ is nonzero only for $i \in [a-2, a]$, the intrinsic alignment contribution to the nulled map $a$ affects exactly the same set of redshift bins. As a result, the nulling property remains valid even in the presence of local intrinsic alignments.}

Our goal in this article is to quantify the constraining power of this test by computing a signal to noise in Sec.~\ref{ssec:SN} and doing sampling tests in Sec.~\ref{ssec:MCMC}. Note, however, that several systematics can affect the nulling procedure and the chosen cross-spectra such that the property $\left<\mathcal{C}^{\BNT{\kappa} {\rm g}}_{ai}(\ell)\right>=0$ might be broken. In Sec.~\ref{sec:bispectrum}, in the context of a Euclid-like survey, we quantify the expected corrections coming from the reduced shear and magnification bias and analyze their effects on our results. We also investigate the effects induced by photometric redshift errors and the gain we could bring to the determination of these in Sec.~\ref{sec:photometric}. Throughout this paper, the correlators are computed in the context of the Limber approximation \cite{Kaiser:1991qi}. We use a matter power spectrum estimated using a Halofit model which is presented in Ref.~\cite{Takahashi:2012em} as a revised version of the model from Ref.~\cite{Smith:2002dz}. Its parameters are set on the best fit of Planck's linear matter power spectrum \cite{Planck:2018nkj}.

\section{Specifications for a Euclid-like survey}\label{sec:specifications}

In this section, we describe the configuration used to mimic a Euclid-like survey. 

\subsection{Redshift distribution of galaxies}\label{ssec:distribution}

We adopt the specifications of the Euclid mission \cite{Euclid:2019clj,Deshpande:2019sdl}. The sky fraction is chosen to be $f_{\rm sky}=0.36$ (corresponding to 15,000 deg$^2$). The redshift distribution of galaxies is described as
\begin{equation}
    n(z)\propto \left(\frac{z}{z_0}\right)^2 \exp \left[-\left(\frac{z}{z_0}\right)^\frac{3}{2}\right],
\end{equation}
where $z_0=z_{\rm m}/\sqrt{2}$ and $z_{\rm m}=0.9$ is the median redshift of the survey. It is normalized so as to get an overall galaxy surface density of $30 {\rm \ arcmin^{-2}}$.

\subsubsection{Source galaxies}\label{sssec:sources}

For the sources, we work with $10$ equally populated redshift bins as in Ref.~\cite{Euclid:2019clj}, each having a galaxy surface density of $\overline{n}_i=3 {\rm \ arcmin^{-2}}$, in the intervals (with sorted $z_i$): $\{0.0010,$ $0.42,$ $0.56,$ $0.68,$ $0.79,$ $0.90,$ $1.02,$ $1.15,$ $1.32,$ $1.57,$ $2.50\}$. We then define the normalized redshift distribution of galaxies in the $i$th bin as
\begin{equation}
n_i(z)=\frac{\int_{z_i}^{z_{i+1}} \dd z_p\, n(z) p_{\rm ph}(z_p,z,\zb{i},\sigma_z)}{\int_{z_{min}}^{z_{max}} \dd z \int_{z_i}^{z_{i+1}} \dd z_p\, n(z) p_{\rm ph}(z_p,z,\zb{i},\sigma_z)},
\end{equation}
where $p_{\rm ph}$ is the probability that a galaxy at redshift $z$ is measured at redshift $z_p$. We choose here a simplified version of the probability from Ref.~\cite{Deshpande:2019sdl} for which we have
\begin{equation}\label{eq:photoprob}
p_{\rm ph}(z_p,z,\zb{i},\sigma_z)= \frac{\exp\left[-\frac{1}{2} \left(\frac{z-z_p-\zb{i}}{\sigma_z (1+z)}\right)^2\right]}
{\sqrt{2 \pi} \sigma_z (1+z)} ,
\end{equation}
where $\sigma_z$ is the variance of the photometric redshift measurement at redshift $0$ and $\zb{i}$ is the mean redshift error in the $i$th redshift bin. Unless stated otherwise, we work with $\sigma_z = 0.02$ which gives distribution shapes that are consistent in width with the ones obtained from the Euclid Flagship simulation \cite{Euclid:2024few}. Notably, we will use for a comparison $\sigma_z=0$—ideal case—and $\sigma_z=0.05$, which is used in Ref.~\cite{Euclid:2019clj} and is consistent with KiDS-1000 \cite{Hildebrandt:2020rno}. In short, $\sigma_z=0.05$ corresponds to optimal photometric precision for Stage III surveys and minimal for Stage IV, and $\sigma_z=0.02$ is an estimate of the optimal photometric precision for Stage IV surveys.

\subsubsection{Low-redshift tracers}\label{sssec:lowzg}

As stated in Sec.~\ref{ssec:crossspectra}, any low-redshift tracers could be used to build the null test, and in an actual implementation, one could use a wide variety of low-redshift tracers including spectroscopic galaxies, photometric galaxies, red-sequence galaxies, galaxy clusters, HI (neutral hydrogen) galaxy maps, radio continuum sources, and intensity maps of the cosmic infrared background or dust emission, to name a few.

Currently, the most natural choice seems to be photometric surveys. In such a nulling test,\footnote{and this is at variance with the shear ratio approach described in Sec.~\ref{sec:shearratio}.} the constraints on the redshift distribution of the foreground galaxies are rather limited as nulling works as soon as galaxies are known to be below a given redshift. On the other hand, the number density of galaxies in photometric surveys is large enough to allow the exploitation of density fluctuations at small physical scales (down to sub-Mpc scales).

Finally, choosing a photometric survey allows us to build a full test while using cross-spectra that a standard $3\times2$pt analysis would produce, except that it will be possible to use much higher multipoles than in standard analysis.

In such a setting, the binned low-redshift galaxy density fields are defined as
\begin{equation}
    \delta^{\rm g}_i (\mathbf{n}) = \int_0^\infty \dd\chi\, n^{\rm g}_i(\chi) \delta_{\rm m}(\chi,\mathbf{n}),
\end{equation}
where $n_i^{\rm g}$ are binned redshift\footnote{One can switch from redshift to comoving distance in galaxy distributions thanks to galaxy number conservation: $\dd \chi\, n(\chi)= \dd z\, n(z)$.} galaxy distributions, which we describe in this section.

One simple solution for low-redshift galaxy distributions is to use the same distributions and binning as for sources. However, when $\sigma_z \neq 0$, this choice is not ideal since the distribution of sources overlap with neighboring bins breaking the nulling property. More precisely, neither the galaxy distribution nor the nulled lensing kernel exactly goes to $0$ for low redshifts when $\sigma_z \neq 0$. To ensure an adequate level of nulling, we could drop some elements of the data vector by having the condition $i\leq a-4$, but this solution reduces the amount of data and therefore degrades performance. A better solution is to choose an optimized binning scheme for low-redshift galaxy distributions.

To do so, we choose different bin limits $z_{\rm g}(i)$, and we have
\begin{equation}\label{eq:lowzn(z)}
n^{\rm g}_i(z)=\frac{\int_{z_{\rm g}(i)}^{z_{\rm g}(i+1)} \dd z_p\, n(z) p_{\rm ph}(z_p,z,\zb{i},\sigma_z)}{\int_{z_{min}}^{z_{max}} \dd z \int_{z_{\rm g}(i)}^{z_{\rm g}(i+1)} \dd z_p\, n(z) p_{\rm ph}(z_p,z,\zb{i},\sigma_z)}.
\end{equation}

The values of $z_{\rm g}(i)$ are chosen such that the nulled lensing kernels with indices $a\geq4$ and the low-redshift galaxy distributions with indices $i\leq a-3$ never overlap if one is higher than $10 \%$ of its maximum value. As such, we ensure that the product of the two is less than $1\%$ of the maximum value it would have for nonvanishing cross-spectra. The detailed methodology along with the values of bin's limits are detailed in Appendix~\ref{sec:ggal}. An example of this condition can be seen in Fig.~\ref{fig:datavector}.

To validate this method, we check the value of the ratio $\mathcal{C}^{\BNT{\kappa} {\rm g}}_{ai}(\ell)/\sqrt{\left<\mathcal{C}^{\BNT{\kappa} {\rm g}}_{ai}(\ell)^2\right>}$ for $i \leq a-3$. The computation of covariances is detailed in Sec.~\ref{ssec:covmat} below. Since we find it to be less than
$1 \% $, we will assume that $\mathcal{C}^{\BNT{\kappa} {\rm g}}_{ai}(\ell)$ is within the noise that is compatible with $\left<\mathcal{C}^{\BNT{\kappa} {\rm g}}_{ai}(\ell)\right> = 0$.

\subsection{Estimating the noise and the covariance matrix}\label{ssec:covmat}

Let us now turn to the computation of the covariance matrix of the data vector. To do so, we must compute a defined set of power spectra, namely,
\begin{align}
\left<\kappa_i(\boldsymbol{\ell})\kappa_j(\boldsymbol{\ell}')\right> &= \mathcal{C}^{\kappa \kappa}_{ij}(\ell)\delta_D(\boldsymbol{\ell}+\boldsymbol{\ell}'),\\
\mathcal{C}^{\BNT{\kappa} \BNT{\kappa}}_{ab}(\ell)&=p_{ai}\mathcal{C}^{\kappa \kappa}_{ij}(\ell)p_{bj},\\
\left<\BNT{\kappa}_a(\boldsymbol{\ell})\delta^{\rm g}_i(\boldsymbol{\ell}')\right> &= \mathcal{C}^{\BNT{\kappa} {\rm g}}_{ai}(\ell)\delta_D(\boldsymbol{\ell}+\boldsymbol{\ell}'),\\
\mathcal{C}^{\BNT{\kappa} {\rm g}}_{ai}(\ell)&=p_{aj}\mathcal{C}^{\kappa {\rm g}}_{ji}(\ell), \\
\left<\kappa_j(\boldsymbol{\ell})\delta^{\rm g}_i(\boldsymbol{\ell}')\right> &= \mathcal{C}^{\kappa {\rm g}}_{ji}(\ell)\delta_D(\boldsymbol{\ell}+\boldsymbol{\ell}'),\\
\left<\delta^{\rm g}_i(\boldsymbol{\ell})\delta^{\rm g}_j(\boldsymbol{\ell}')\right> &= \mathcal{C}^{\rm gg}_i(\ell)\delta_D(\boldsymbol{\ell}+\boldsymbol{\ell}')\delta_{ij}.
\end{align}

Let us now define the noise in these spectra following \cite{Euclid:2019clj,Deshpande:2019sdl}. For $\mathcal{C}^{\BNT{\kappa} \BNT{\kappa}}_{ab}$, the shape noise reads $\BNT{\mathcal{S}}_{ab}=p_{ai}  \mathcal{S}_{ij} p_{bj}$ where $\mathcal{S}_{ij}=\delta_{ij} \sigma_{\rm s}^2/\overline{n}_i$ and we expect to have $\sigma_{\rm s}=0.3$. For $\mathcal{C}^{\rm gg}_{i}$, the galaxy shot noise is given by $\mathcal{N}^{\rm gg}_{ij}=\delta_{ij}/\overline{n}^{\rm g}_i$. $\overline{n}_i$ is the total number of galaxies in bin $i$ with g indicating the low-redshift galaxy bins. 

We are now able to compute the correlation matrix, assuming the classical Knox approximation \cite{Knox:1995dq}, for indices $ai$ such that $1 \leq i \leq a-3 \leq 7$ ($28 \times 28$ matrix),
\begin{equation}\label{eq:covmat}
\Sigma_{ai,bj}(\ell)\!=\!\frac{\delta_{i,j}}{(2\ell+1)f_{\rm sky}} \left( \mathcal{C}^{\BNT{\kappa} \BNT{\kappa}}_{ab}(\ell)+\BNT{\mathcal{S}}_{ab}\right) \left( \mathcal{C}^{\rm gg}_i(\ell)+\mathcal{N}^{\rm gg}_i \right).
\end{equation}

This expression is a direct consequence of the independence of the fields $\BNT{\kappa}_a(\mathbf{n})$ and $\delta^{\rm g}_i(\mathbf{n})$. We also neglected terms including nulled cross-spectra as these are expected to vanish. Indeed, one can check that if $i \leq a-3$ and $j \leq b-3$ then we necessarily have $j \leq a-3$ or $i \leq b-3$ such that either $\mathcal{C}^{\BNT{\kappa} {\rm g}}_{aj}$ or $\mathcal{C}^{\BNT{\kappa} {\rm g}}_{bi}$ is nulled. The trispectrum term is also negligible as, for the same reasons on the indices, the four kernels never overlap altogether. In the end, the covariance matrix is not affected by the supersample covariance  nor by connected non-Gaussianities. As a result, the covariance matrix can be evaluated directly from measurements that would already be estimated for the survey. This is only valid in the Limber approximation, and it is done at the cost of noisy covariance. However, it makes the constraints independent of any modeling of the matter distribution properties such as the matter power spectra $P(k)$. Note also that the test is independent of the precision with which the redshifts of the nearby galaxies are known, provided it is in the range in which nulling is expected. This is a clear difference from the so-called shear ratio test as described below.

\subsection{Methodology comparison with shear ratio}
\label{sec:shearratio}
Before turning to the study of systematics that might affect nulling as a cosmological probe, let us first describe how our nulling approach compares to similar ideas in the literature. Indeed, some nulling properties have already been exploited in recent large-scale surveys such as the Dark Energy Survey (DES) but with a different approach, as we will describe here. Notably, the use of the so-called shear ratio in Refs.~\cite{DES:2018lpj,DES:2021jzg} can be identified with the local nulling techniques described in Ref.~\cite{Joachimi:2010va}. The reasoning is as follows: if the low-redshift galaxy distributions are narrow enough, then the ratios
\begin{equation}\label{eq:lsr}
    q^{i,j,k}(\ell)=\frac{\mathcal{C}^{\kappa {\rm g}}_{ji}(\ell)}{\mathcal{C}^{\kappa {\rm g}}_{ki}(\ell)}
\end{equation}
are constant and depend only on the background geometry of space-time. Although DES uses shear ratios as observables, one could rewrite it in a linearized version and compare
\begin{equation}
    \mathcal{C}^{\kappa {\rm g}}_{ji}(\ell)-q^{i,j,k}\mathcal{C}^{\kappa {\rm g}}_{ki}(\ell)
\end{equation}
to zero where $q^{i,j,k}$ would be computed from some assumed background geometry. This gives a local nulling which means that the combination of cross-spectra that induce nulling depends on the chosen low-redshift galaxy sample. Moreover, it only works for the galaxy-galaxy lensing two-point cross-spectra and with the condition that this low-redshift galaxy sample is small enough to be identified to a plane. 
The BNT transform, on the other hand, gives an exact and global nulling as no assumption has to be made on samples, and the nulling property is satisfied on all low redshifts at the level of the shear maps themselves. In other words, the BNT transform exhibits an internal symmetry of shear maps. Thus, one can use any low-redshift tracer with few assumptions on its position and shape, and one can even use alternative observables or higher-order statistics. One straightforward example could be weak-lensing nulled spectra $\left<\BNT{\kappa}_a(\boldsymbol{\ell})\BNT{\kappa}_b(\boldsymbol{\ell}')\right>$ with well-chosen indices.

However, one can note similarities in the features from these two methods as both provide us with an observable that is independent of the matter power spectrum and depends on the background geometry (excluding $H_0$). Moreover, let us also point to the fact that lensing shear ratios carry the same theoretical degeneracy as the BNT transform described by Eq.~\eqref{eq:degeneracy}. However, it does not concern the same parameter space as, for lensing shear ratio, the relation from Eq.~\eqref{eq:chiz} is probed for both the sources and low-redshift galaxies, implying that the redshifts that are probed are different.

To summarize, the two methods probe the same relation at different redshifts and with different methods and prerequisites, meaning that they exploit the data differently and so they access different information content. The details of our theoretical investigation of lensing shear ratio observable are provided in Appendix~\ref{sec:lsr}.

\section{Bispectrum bias}\label{sec:bispectrum}

Besides systematic effects identified above, nulling is ensured as long as Eq.~\eqref{eq:convergence} is valid, that is, as long as correction terms to the linear relation between observed shear and local density is preserved.

However, departures from this regime can occur at small scales, when approaching the strong lensing regime, that is, when we do not have $\kappa \ll 1$. In that respect, the leading corrections are the reduced shear effects and the impact of the magnification bias; see Ref.~\cite{Deshpande:2019sdl}. In this section, we investigate these corrections at bispectrum level.

\subsection{Reduced shear}

Cosmic shear and convergence are not direct observables. Instead, measuring the ellipticities of galaxies gives us access to the reduced shear, which is defined as
\begin{equation}\label{eq:RS}
    g^\alpha(z_{\rm S},\mathbf{n})=\frac{\gamma^\alpha(z_{\rm S},\mathbf{n})}{1-\kappa(z_{\rm S},\mathbf{n})},
\end{equation}
where $\mathbf{n}$ is the position of the source galaxy on the sky, $z_{\rm S}$ the redshift of the source, $\gamma$ the shear field, $\kappa$ the convergence field, and $\alpha=1$ or $2$ denote the two components of the shear.
Nulling can then be applied to the reduced shear in bin $i$, with source distribution $n_i(z_{\rm S})$, yielding
\begin{equation}
    \hat{g}^\alpha_i(\mathbf{n})=p_{ij}g^\alpha_j(\mathbf{n}),
\end{equation}
where
\begin{equation}\label{eq:binnedRS}
    g^\alpha_j(\mathbf{n})=\int \dd z_{\rm S}\, n_j(z_{\rm S})g^\alpha(z_{\rm S},\mathbf{n})
\end{equation}
is the reduced shear in bin $j$ before nulling.

In Eq.~\eqref{eq:RS}, the factor $1-\kappa$ induces a correction on the shear and consequently a departure from the nulling property itself. We must estimate this error and try to correct for it if necessary.

As $|\gamma| \ll 1$ and $|\kappa| \ll 1$, we can expand the nulled reduced shear field in a perturbative manner such that
\begin{equation}
    \hat{g}^\alpha_i(\mathbf{n})\approx\hat{\gamma}^\alpha_i(\mathbf{n})+\sum_j p_{ij} \int \dd z_{\rm S}\, n_j(z_{\rm S})\gamma^\alpha(z_{\rm S},\mathbf{n})\kappa(z_{\rm S},\mathbf{n}).
\end{equation}

We then follow the formalism recalled in Ref.~\cite{Deshpande:2019sdl} to use the E-mode of the reduced shear field in bin $i$. It is expressed in Fourier space as
\begin{equation}
    E_i(\boldsymbol{\ell})=T^1(\boldsymbol{\ell})g^1_i(\boldsymbol{\ell})+T^2(\boldsymbol{\ell})g^2_i(\boldsymbol{\ell}),
\end{equation}
where $\boldsymbol{\ell}$ is the spherical harmonic conjugate of $\mathbf{n}$, $T^1(\boldsymbol{\ell})=\cos(2\phi_\ell)=\cos^2(\phi_\ell)-\sin^2(\phi_\ell)=\frac{\ell^2_x-\ell^2_y}{\ell^2}$, $T^2(\boldsymbol{\ell})=\sin(2\phi_\ell)=2\cos(\phi_\ell)\sin(\phi_\ell)=2\frac{\ell_x\ell_y}{\ell^2}$, and $\phi_\ell$ is the angular component of vector $\boldsymbol{\ell}$ which has magnitude $\ell$. Moreover, in the "prefactor unity approximation" \citep{Kitching:2016zkn}, which effect is negligible for a Euclid-like survey \citep{Kilbinger:2017lvu}, let us recall that the shear $\gamma_i$ can be determined from the convergence $\kappa_i$ as
\begin{equation}
    \gamma^\alpha_i(\boldsymbol{\ell})=T^\alpha(\boldsymbol{\ell})\kappa_i(\boldsymbol{\ell}).
\end{equation}

\subsection{Magnification bias}

The magnification bias is also known to affect both the observed galaxy distribution and shear as it increases (respectively, decreases) the observed number density of galaxies in magnified (respectively, demagnified) regions. We follow the formalism recalled in Ref.~\cite{Deshpande:2019sdl} such that
\begin{align}
\delta^{\rm g}_{obs,i}(\mathbf{n})&=\delta^{\rm g}_i(\mathbf{n})+(5s_i^{\rm g}-2)\kappa^{\rm g}_i(\mathbf{n}),\\
\delta^{s}_{obs,i}(\mathbf{n})&=\delta^{s}_i(\mathbf{n})+(5s_i-2)\kappa_i(\mathbf{n}),\\
\gamma^\alpha_{obs,i}(\mathbf{n})&=\gamma^\alpha_i(\mathbf{n})(1+\delta^{s}_{obs,i}(\mathbf{n})),
\end{align}
where $\delta_i$ are binned galaxy density fields, $\kappa_i$ are binned convergence fields, $\gamma_i^\alpha$ are binned shear fields, "obs" stands for observed (i.e. corrected) fields
and we use different indices for the low-redshift galaxy "g" and source "s" distributions since the binning can be different as described in Sec.~\ref{ssec:distribution}. Finally, $s_i$ is the slope of the cumulative galaxy number counts above the magnitude limit of the survey in redshift bin $i$ given by
\begin{equation}
    s_i=\left.\frac{\dd \log n(\overline{z}_i,m)}{\dd m}\right|_{m=m_{\rm lim}}
\end{equation}
and which we determine following Euclid's Third Science Performance Verification \citep{Euclid:2024few,Euclid:2024yrr} according to the fitting formula given in Appendix~\ref{sec:MBSlope}.

In particular, one should pay attention to the binning scheme, as any product of fields should be binned after the product, such that, for instance,
\begin{equation}
    \gamma^\alpha_{obs,i}(\mathbf{n})=\gamma^\alpha_i(\mathbf{n})+(\gamma^\alpha \delta^{s})_i(\mathbf{n})+(5s_i-2)(\gamma^\alpha \kappa)_i(\mathbf{n}),
\end{equation}
where $(\gamma^\alpha \kappa)_i$ is computed in the same way as in Eq.~\eqref{eq:binnedRS} and
\begin{equation}
    (\gamma^\alpha \delta^{s})_i(\mathbf{n})=\int \dd z_{\rm S}\, n_i(z_{\rm S}) \gamma^\alpha(z_{\rm S},\mathbf{n}) \delta_{\rm m}(z_{\rm S},\mathbf{n}).
\end{equation}

\subsection{Bispectrum corrections}\label{ssec:corr}

We can now compute the correction from these effects to the cross-spectra $\mathcal{C}^{\BNT{\kappa} {\rm g}}_{ai}$. We work within the Limber approximation, which somehow decorrelates planes whose redshifts are distant from one another. As a consequence, it makes $(\gamma^\alpha \delta^{s})_i$ and all second-order terms in the deformation matrix (lens-lens correlations and geodesic deviations) vanish. With the convolution theorem, we then find
\begin{align*}
&\left<\delta^{\rm g}_{obs,i}(\boldsymbol{\ell})E_{obs,j}(\boldsymbol{\ell}')\right>=\left<\delta^{\rm g}_i(\boldsymbol{\ell})\kappa_j(\boldsymbol{\ell}')\right> \\
&+(5s_i^{\rm g}-2)\left<\kappa^{\rm g}_i(\boldsymbol{\ell})\kappa_j(\boldsymbol{\ell}')\right>\\ 
&+\int \frac{d^2\ell_1}{\sqrt{2 \pi}^2}\cos(2\phi_{\ell_1}-2\phi_\ell) (5s_j-1) \int \dd z_{\rm S}\, n_j(z_{\rm S}) \\
&\left<[\delta^{\rm g}_i(\boldsymbol{\ell})+(5s_i^{\rm g}-2)\kappa^{\rm g}_i(\boldsymbol{\ell})]\kappa(z_{\rm S},\boldsymbol{\ell}_1) \kappa(z_{\rm S},\boldsymbol{\ell}'-\boldsymbol{\ell}_1)\right>.
\end{align*}

We can then separate the contributions between the original signal, terms involving the matter power spectrum, and terms involving the bispectrum. Let us now turn to the nulled cross-spectra
\begin{equation}
    \left<\delta^{\rm g}_{obs,i}(\boldsymbol{\ell})\hat{E}_{obs,a}(\boldsymbol{\ell}')\right>=\left[\mathcal{C}^{\BNT{\kappa} {\rm g}}_{ai}(\boldsymbol{\ell})\!+\!\delta_P \mathcal{C}^{\BNT{\kappa} {\rm g}}_{ai}(\boldsymbol{\ell}) \!+\! \delta_B \mathcal{C}^{\BNT{\kappa} {\rm g}}_{ai}(\boldsymbol{\ell})\right]\delta_D(\boldsymbol{\ell}+\boldsymbol{\ell}'),\nonumber
\end{equation}
where
\begin{equation}
    \delta_P \mathcal{C}^{\BNT{\kappa} {\rm g}}_{ai}(\ell) =(5s_i^{\rm g}-2)\!\!\!\int\!\! \frac{\dd\chi}{\chi^2} \!\!\int \dd z_{\rm S}' n_i^{\rm g}(z_{\rm S}')
    w(z_{\rm S}',\chi)\BNT{w}_a(\chi)P\left(\frac{\ell}{\chi},\chi\right),\nonumber
\end{equation}
and
\begin{align*}
    &\delta_B \mathcal{C}^{\BNT{\kappa} {\rm g}}_{ai}(\ell)= \sum_j p_{a,j}\int \dd z_{\rm S} n_j(z_{\rm S}) \int \frac{\dd\chi}{\chi^4} (5s_j-1) \\
    &\left(n_i^{\rm g}(\chi)+(5s_i^{\rm g}-2)\int \dd z_{\rm S}'  n_i^{\rm g}(z_{\rm S}')w(z_{\rm S}',\chi)\right) \\
    & w(z_{\rm S},\chi)^2 \! \int  \! \frac{\dd^2 \ell_1}{(2 \pi)^2} \cos(2\phi_{\ell_1}-2\phi_\ell) B\left(\frac{\boldsymbol{\ell}}{\chi},\frac{\boldsymbol{\ell}_1}{\chi},\frac{-\boldsymbol{\ell}-\boldsymbol{\ell}_1}{\chi}\right).
\end{align*}

Because of the nulling property and the choice $i\leq a-3$, $\delta_P \mathcal{C}^{\BNT{\kappa} {\rm g}}_{ai}(\ell)$ will be below the noise level as the lensing kernels of $\kappa^{\rm g}_i$ and $\BNT{\kappa}_a$ do not overlap. However, it is not the case for $\delta_B \mathcal{C}^{\BNT{\kappa} {\rm g}}_{ai}(\ell)$ due the structure of $n_j(z_{\rm S})(5s_j-1)w(z_{\rm S},\chi)^2$ which happens to have a large support. We will see how to correct for this effect in Sec.~\ref{ssec:bispectrum} below.

\subsection{Estimating the bispectrum}\label{ssec:bispectrum}

To estimate the correction described in Sec.~\ref{ssec:corr}, one needs in particular to model the bispectrum. We choose here to follow the fitting formula from Ref.~\cite{Scoccimarro:2000ee},
\begin{equation}
    B(\boldsymbol{k}_1,\boldsymbol{k}_2,\boldsymbol{k}_3,\chi)=2F_2^{\rm eff}(\boldsymbol{k}_1,\boldsymbol{k}_2) P(k_1,\chi)P(k_2,\chi) + {\rm cyc}.
\end{equation}
where the effective kernel $F^{\rm eff}_2$ is given by
\begin{align}
F_2^{\rm eff}(\boldsymbol{k}_1,\boldsymbol{k}_2)&=\frac{5}{7} a(n,k_1)a(n,k_2) \nonumber\\
&+ \frac{1}{2} \boldsymbol{k}_1\cdot\boldsymbol{k}_2 \left( \frac{1}{k_1^2} + \frac{1}{k_2^2} \right) b(n,k_1)b(n,k_2) \nonumber\\
&+ \frac{2}{7} \left(\frac{\boldsymbol{k}_1\cdot\boldsymbol{k}_2}{k_1 k_2}\right)^2 c(n,k_1)c(n,k_2)\nonumber,
\end{align}
and $n_{\rm s}$ is the spectral index. As demonstrated in Ref.~\cite{Scoccimarro:2000ee}, at small enough scales, the function $a$ does not depend on the wave vector and $b = c = 0$. This eventually yields
\begin{equation}\label{eq:bispectrum}
    B(\boldsymbol{k}_1,\boldsymbol{k}_2,\boldsymbol{k}_3,\chi)=Q_3 P(k_1,\chi)P(k_2,\chi) + {\rm cyc}.
\end{equation}

To take into account the bias created by the bispectrum corrections, we simply add the bispectrum normalization $Q_3$ to the parameter space and will eventually have to marginalize over it.

\section{Cosmological constraints from nulling}\label{sec:constraints}

In this section, we will describe our numerical results in terms of cosmological constraints. We first choose the data vector to be $\{\observed{\mathcal{C}}^{\BNT{\kappa}, {\rm g}}_{(ai)}(\ell)\}$ where we denote observed quantities as $\observed{X}$, for $a \in \{4,\dots,N_z=10\}$ and $i \in \{1,\dots,a-3\}$ (28 components) and 32 bins of wave modes $\ell$, logarithmically spaced within $\{10,\dots,10^{4.2}\approx 16\,000\}$. As $\observed{\mathcal{C}}^{\kappa {\rm g}}_{ji}(\ell)$ is measured, we will only vary the nulling coefficient $p_{aj}(\boldsymbol{\theta})$ such that $\observed{\mathcal{C}}^{\BNT{\kappa} {\rm g}}_{ai}(\ell)=p_{aj}(\boldsymbol{\theta})\observed{\mathcal{C}}^{\kappa {\rm g}}_{ji}(\ell)$. For the sake of clarity, while simplifying the notations in the following, let us define a derivative operator ${\rm D}_\theta$ as ${\rm D}^n_\theta \mathcal{C}^{\BNT{\kappa} {\rm g}}_{ai}(\ell)=\frac{\partial^n p_{aj}}{\partial \theta^n} \times \mathcal{C}^{\kappa {\rm g}}_{ji}(\ell)$. Hence, if the parameters used to perform the nulling correspond to the right underlying cosmology and without systematics, then the resulting signal should be statistically compatible with zero, which gives the condition
\begin{equation}
    \langle \observed{\mathcal{C}}^{\BNT{\kappa} {\rm g}}_{ai}(\ell) \rangle = 0,\ \mathrm{for\ all}\ \ell \ \mathrm{and}\ \mathrm{for}\ i \leq a-3.
\end{equation}
This condition therefore provides us in principle with constraints on the parameters. For a Euclid-like survey, we could have $\boldsymbol{\theta} \in \{\Omega_{\rm m},\Omega_K,{\rm w}_0,{\rm w}_a\}$ with fiducial values $\{0.32,0,-1,0\}$. However, in this study, we will focus on $\{\Omega_{\rm m},{\rm w}_0\}$, in flat universe, but we will add $Q_3$ as defined in Eq.~\eqref{eq:bispectrum}, with fiducial value $1$, to take bispectrum bias into account. The nulling condition then becomes
\begin{equation}
    \langle \observed{\mathcal{C}}^{\BNT{\kappa} {\rm g}}_{ai}(\ell) \rangle = Q_3 \times \delta_B \mathcal{C}^{\BNT{\kappa} {\rm g}}_{ai}(\ell),
\end{equation}
where $\delta_B \mathcal{C}^{\BNT{\kappa} {\rm g}}_{ai}(\ell)$ is computed for fiducial parameters and used as a template.

\subsection{Signal and noise modeling}\label{ssec:SN}

To quantify our ability to gain information from nulling, we shall define a signal and noise. Hence, let us define an effective amplitude of the induced signal for a variation of parameter $\delta \theta$ as
\begin{equation}\label{eq:dchi}
\delta \mathcal{C}^{\BNT{\kappa} {\rm g}}(\ell)=\delta \theta \left[n_b(N_z) \sum_{(ai)} \left( {\rm D}_\theta \mathcal{C}^{\BNT{\kappa} {\rm g}}_{(ai)}(\ell)\right)^2\right]^\frac{1}{2},
\end{equation}
where $n_b(N_z)$ is the number of components of the data vector (it makes $\delta \mathcal{C}^{\BNT{\kappa} {\rm g}}$ roughly independent of $N_z$). Using the inverse covariance matrix $\Sigma^{-1}_{(ai),(bj)}(\ell)$, we can now compute the $\chi^2$ value of an event corresponding to an observed data vector $\mathcal{C}^{\BNT{\kappa} {\rm g}}_{(ai)}(\ell)$ as
\begin{equation}
\mathcal{R}^{\rm model}(\ell)=\frac{1}{2} \sum_{(ai),(bj)} \Sigma^{-1}_{(ai),(bj)}(\ell) \mathcal{C}^{\BNT{\kappa} {\rm g}}_{(ai)}(\ell) \mathcal{C}^{\BNT{\kappa} {\rm g}}_{(bj)}(\ell),
\end{equation}
where the superscript "model" indicates the parameters used to compute the BNT transform. To be able to visualize the signal compared to a noise amplitude, we also define a noise as
\begin{equation}
\hat{N}(\ell)= \left. \frac{\delta \mathcal{C}^{\BNT{\kappa} {\rm g}}(\ell)}{\delta \theta} \middle/ \left({\rm D}^2_\theta \mathcal{R}^{\rm model}(\ell)\right)^\frac{1}{2} \right. .
\end{equation}

\begin{figure}[!ht]
   \begin{center}
    \includegraphics[width=\columnwidth]{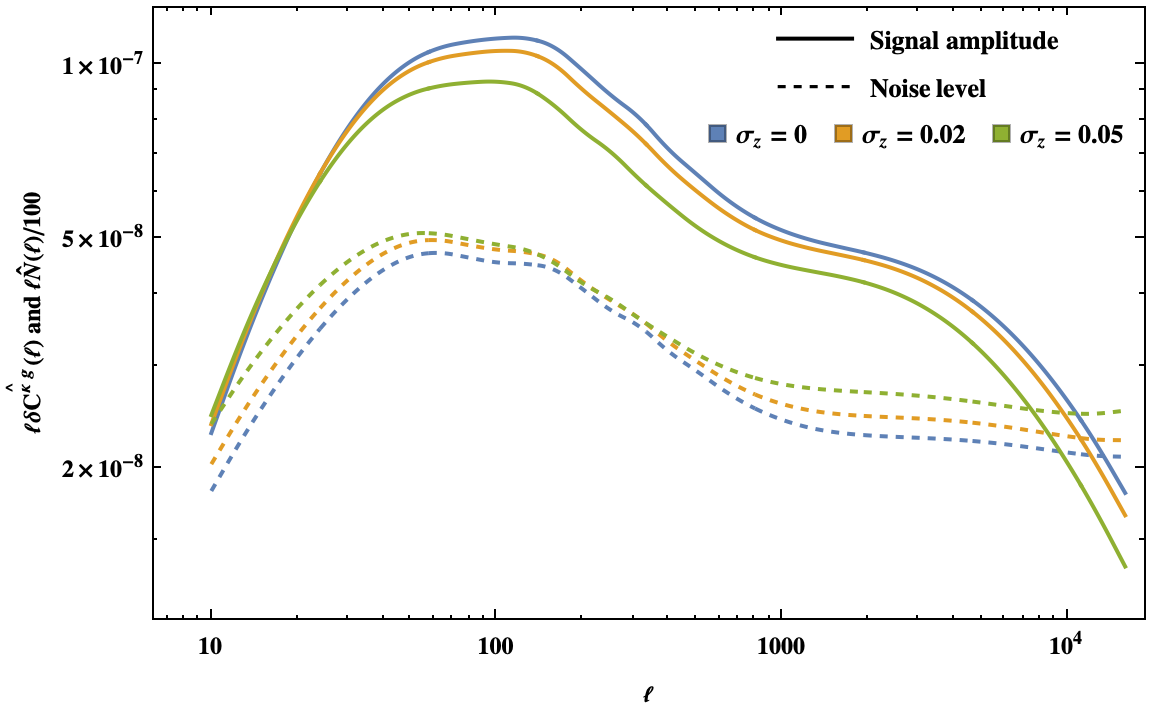}
  \end{center}
  \caption{Amplitude of the signal $\ell \delta \mathcal{C}^{\BNT{\kappa} {\rm g}}(\ell)$ for a $10\%$ variation of ${\rm w}_0$ and noise level $\ell \hat{N} (\ell)/100$ for various $\sigma_z$ values.}
  \label{fig:signalnoisew0}
\end{figure}

We show as an example in Fig.~\ref{fig:signalnoisew0} the signal $\ell \delta \mathcal{C}^{\BNT{\kappa} {\rm g}}(\ell)$ and noise $\ell \hat{N}(\ell)$ amplitudes for a variation of $10\%$ of ${\rm w}_0$. The noise is divided by $100=\sqrt{10,000}$ as roughly $10,000 \ \ell$ modes are available. For visual purposes, we multiplied the signal and noise by $\ell$ to avoid a decrease in the amplitudes with $\ell$. In this example, we see that the signal is $1$ order of magnitude higher than the noise, and we can then safely conclude that we are able to reject models with $10\%$ variation of ${\rm w}_0$.

\begin{figure}[!ht]
   \begin{center}
    \includegraphics[width=\columnwidth]{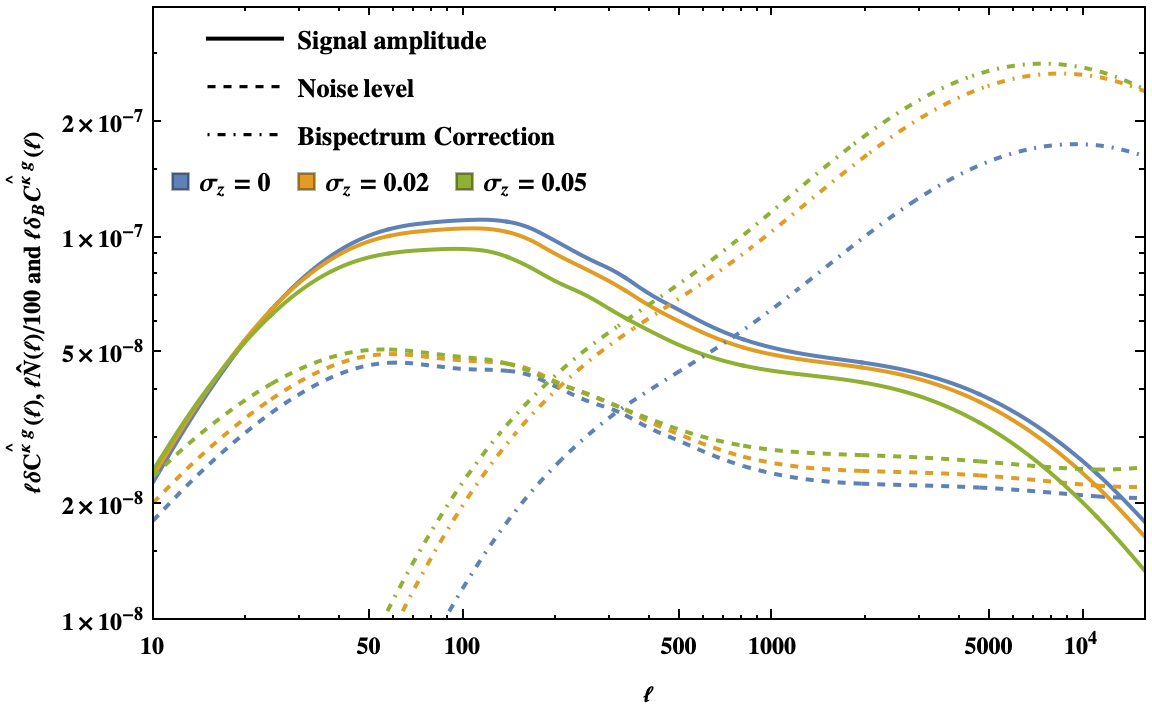}
  \end{center}
  \caption{Amplitude of the signal $\ell \delta \mathcal{C}^{\BNT{\kappa} {\rm g}}(\ell)$ for a $10\%$ variation of ${\rm w}_0$, noise level $\ell \hat{N} (\ell)/100$, and bispectrum correction $\ell \delta_B \mathcal{C}^{\BNT{\kappa} {\rm g}}(\ell)$ for various $\sigma_z$ values.}
  \label{fig:signalnoisew0B}
\end{figure}

To estimate the effect of bispectrum corrections, in Fig.~\ref{fig:signalnoisew0B}, we added with dotted lines the effective amplitude of the vector $\delta_B \mathcal{C}^{\BNT{\kappa} {\rm g}}_{(ai)}(\ell)$ for $Q_3=1$. This shows that, in order to take enough scales into account, we have no choice but to correct for the bispectrum bias by adding another parameter.

Then, one can encode the precision with which two parameters, $\theta$ and $\lambda$, can be measured simultaneously in the Fisher matrix given by
\begin{align}\label{eq:FoM}
F^{\rm model}_{\theta \lambda} &= \sum_\ell {\rm D}_\theta{\rm D}_\lambda \mathcal{R}^{\rm model}(\ell) \\
&= \sum_{l,(ai),(bj)}  \Sigma^{-1}_{(ai),(bj)}(\ell) {\rm D}_\theta\mathcal{C}^{\BNT{\kappa} {\rm g}}_{(ai)}(\ell) {\rm D}_\lambda\mathcal{C}^{\BNT{\kappa} {\rm g}}_{(bj)}(\ell) .\nonumber
\end{align}
The inverse of $F_{\theta \lambda}$ gives the covariance matrix of the parameters, to first order in variations of the data vector. After marginalizing over all other parameters and restricting the analysis to two cosmological parameters, the figure of merit (FOM) is defined as the inverse of the determinant of the resulting covariance matrix. We will not further investigate the Fisher Matrix for now as the cosmological constraints are much too non-gaussian. Thus, we will now sample the parameter space.

\subsection{Monte Carlo Markov chain}\label{ssec:MCMC}

Since the computation of the nulling coefficients is fast, it is easy to sample the parameter space. In this section, we will show results obtained with a Monte Carlo Markov chain (MCMC) sample using Metropolis-Hastings with the help of Cobaya \cite{Torrado:2020dgo,2019ascl.soft10019T}.

We define a Gaussian likelihood as
\begin{align}\label{eq:loglikelihood}
    &\log\mathcal{L}(\boldsymbol{\theta},Q_3) = \\ 
    &-\frac{1}{2} \sum_\ell (\observed{\mathcal{C}}^{\BNT{\kappa} {\rm g}}-Q_3 \delta_B \mathcal{C}^{\BNT{\kappa} {\rm g}})^T \cdot \Sigma^{-1} \cdot (\observed{\mathcal{C}}^{\BNT{\kappa} {\rm g}}-Q_3 \delta_B \mathcal{C}^{\BNT{\kappa} {\rm g}}), \nonumber
\end{align}
where we ignored the $\ell$ dependency and used a matrix notation avoiding $ai$ indices in the equation to make it more compact. $\observed{\mathcal{C}}^{\kappa {\rm g}}_{ji}(\ell)$ is the measured two-point cross correlator, including the magnification bias and reduced shear. We also ignored all explicit $\boldsymbol{\theta}$ dependence in the equation to make it more compact, but both the observed cross spectra $\observed{\mathcal{C}}^{\BNT{\kappa} {\rm g}}_{ai}(\ell)$ and higher-order bias correction $\delta_B \mathcal{C}^{\BNT{\kappa} {\rm g}}_{ai}(\ell)$ do depend on the cosmological parameters as they are BNT transformed quantities. We checked that the corrections of the covariance matrix for bispectrum effects and ignored nulled data are negligible.

\begin{figure}[!ht]
   \begin{center}
    \includegraphics[width=\columnwidth]{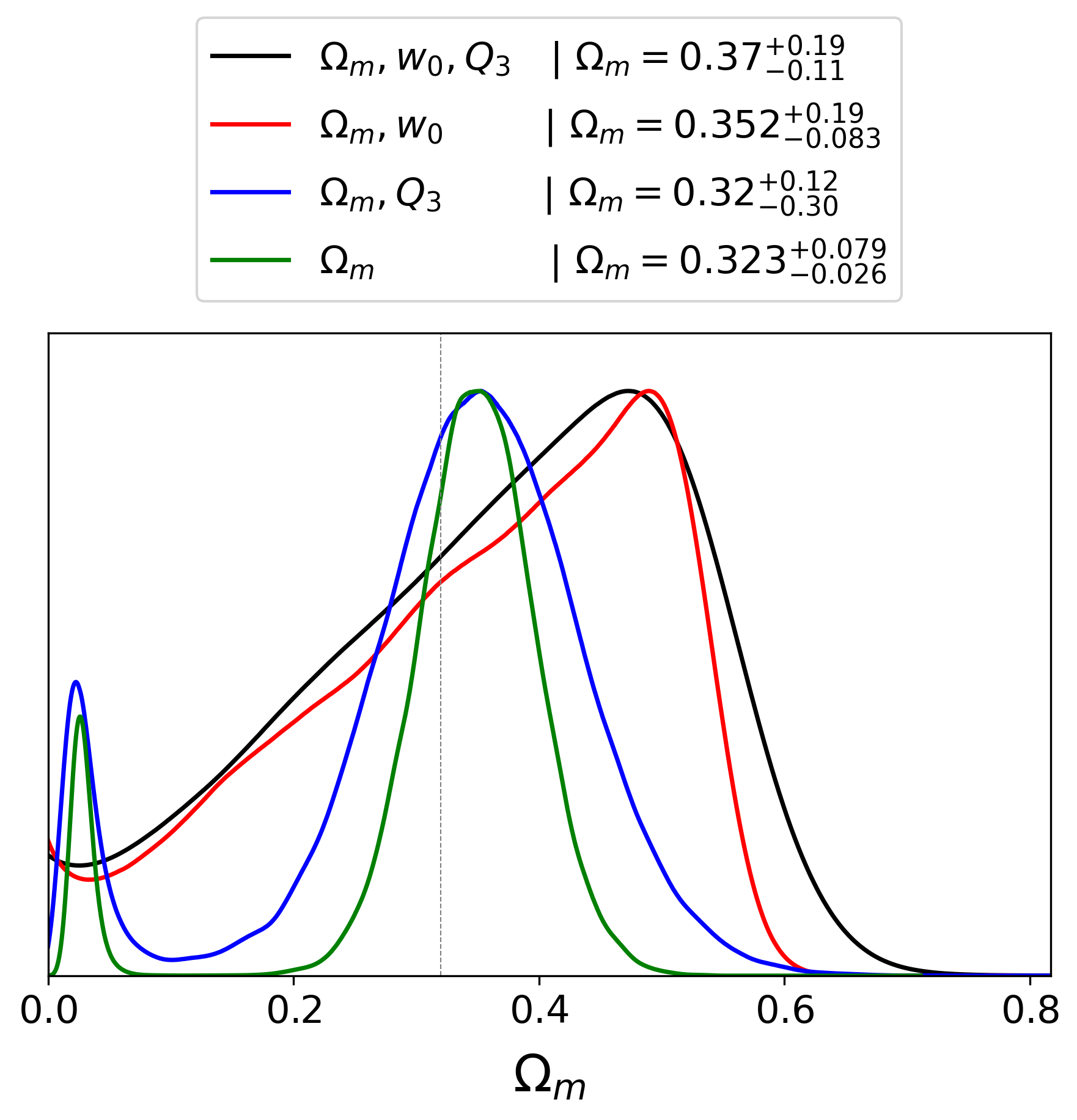}
  \end{center}
  \caption{Posterior distribution for $\Omega_{\rm m}$ with sampled parameters $\{\Omega_{\rm m},{\rm w}_0,Q_3\}$. We show different cases of fixed parameters. In green, $\{{\rm w}_0,Q_3\}$ are fixed. In blue, ${\rm w}_0$ is fixed. In red, $Q_3$ is fixed.}
  \label{fig:MCMC1d}
\end{figure}

As a first step, we estimate the confidence intervals and posteriors obtained for $\Omega_{\rm m}$ in various situations where we sample on $\{\Omega_{\rm m},{\rm w}_0,Q_3\}$ but fix some of the parameters. We show the posterior along with confidence intervals in Fig.~\ref{fig:MCMC1d}.

\begin{figure}[!ht]
   \begin{center}
    \includegraphics[width=\columnwidth]{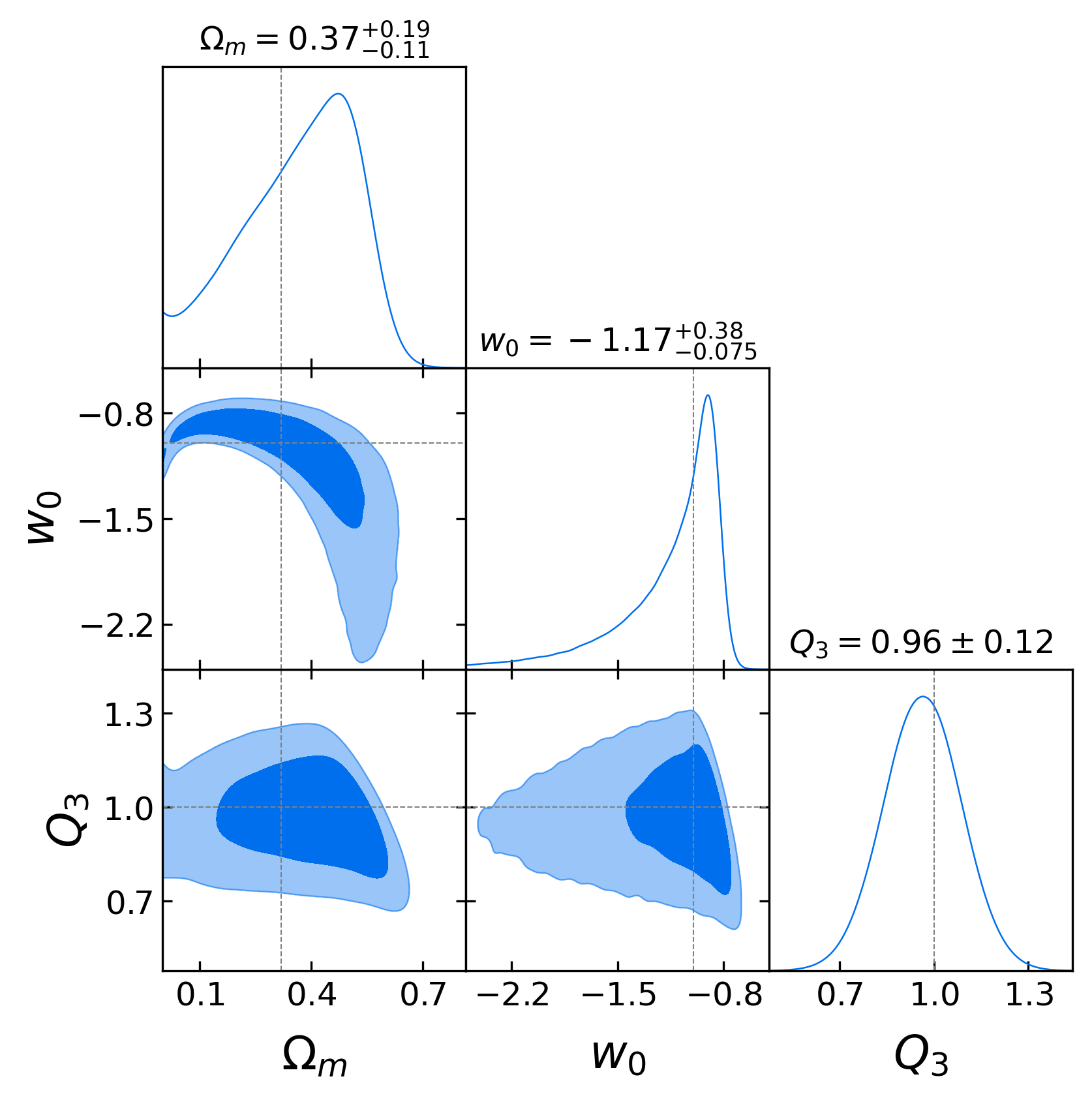}
  \end{center}
  \caption{Triangle plot for sampled parameters $\{\Omega_{\rm m},{\rm w}_0,Q_3\}$. Dashed lines show the fiducial values of parameters.}
  \label{fig:MCMC}
\end{figure}

We see here the effect that adding ${\rm w}_0$ to the parameters can have. It is again an illustration of the degeneracy we showed in Eq.~\eqref{eq:degeneracy} and which we will discuss further later in this section. Figure~\ref{fig:MCMC1d} also illustrates the expected precision loss due to the bispectrum corrections when adding $Q_3$. One can also see a local optimum for low $\Omega_{\rm m}$ when we do not sample for ${\rm w}_0$. This feature arises because we are effectively slicing through a higher-dimensional posterior distribution that has a curved, “banana-shaped” geometry as shown in Fig.~\ref{fig:MCMC}. When ${\rm w}_0$ is held fixed, we are examining a one-dimensional cut through this surface, which can intersect regions of locally higher probability density—even if they do not correspond to the global maximum. This results in an artificial secondary peak at low $\Omega_{\rm m}$. Once ${\rm w}_0$ is included in the sampled parameters, the full degeneracy structure is accounted for and the spurious peak disappears. However, we still observe a slight bump as $\Omega_{\rm m} \to 0$, now due to projection effects rather than slicing. This residual feature has limited practical importance, as values $\Omega_{\rm m} < 0.1$ are firmly excluded by current observations and could have been ruled out directly through the choice of priors.

\begin{figure}[!ht]
   \begin{center}
    \includegraphics[width=\columnwidth]{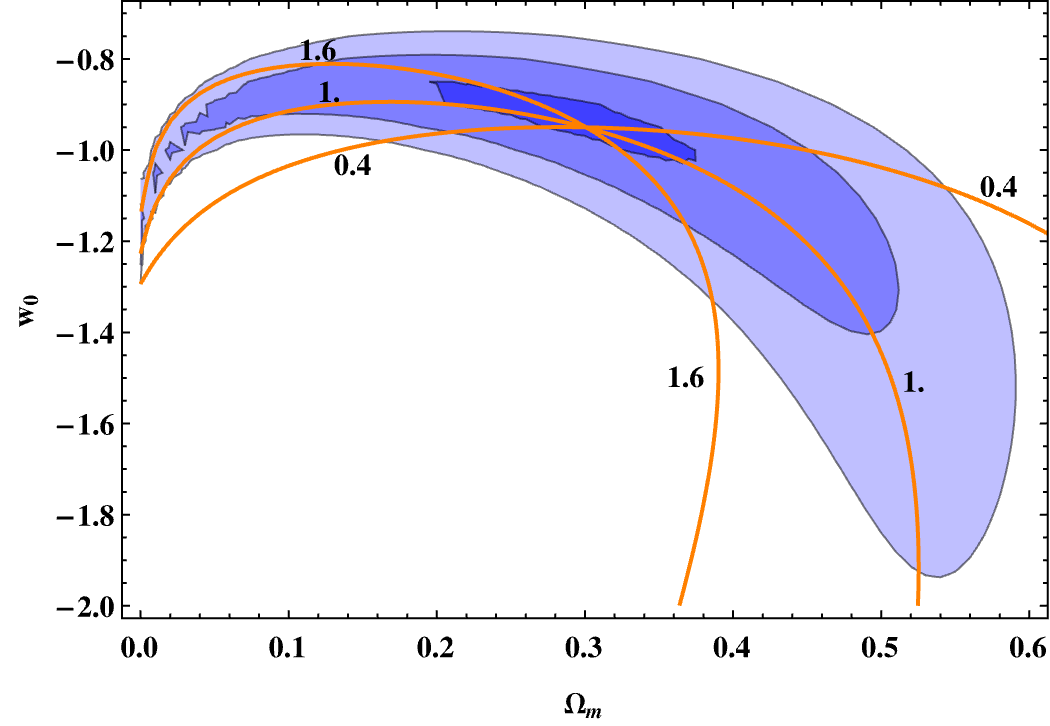}
  \end{center}
  \caption{$\chi^2$ contour plot for $\{\Omega_{\rm m}$, ${\rm w}_0\}$ marginalized over $Q_3$ in blue. Degeneracy lines at fixed redshifts, indicated next to the curves, in orange. The reference model for these lines is $\{\Omega_{\rm m}=0.30,{\rm w}_0=-0.95\}$.}
  \label{fig:degeneracy}
\end{figure}

We show in Fig.~\ref{fig:MCMC} the triangle plot obtained for $\Omega_{\rm m}$ and ${\rm w}_0$ when sampling the three parameters $\{\Omega_{\rm m},{\rm w}_0,Q_3\}$.
As expected, we find a correlation between the two parameters $\Omega_{\rm m}$ and ${\rm w}_0$, pinpointing the need to combine this method with other usual weak lensing and galaxy clustering probes. However, let us emphasize that the nulling test is still adding constraints on the parameters as this correlation is not the same as in usual weak lensing analysis. The contour is indeed perpendicular to the usual constraints such as those found in Ref.~\cite{Euclid:2024yrr}. Let us now determine the origin of this correlation. Since we suspect that it may correspond to the degeneracy identified in Eq.~\eqref{eq:degeneracy}, we try to characterize this degeneracy in the parameter space $\{\Omega_{\rm m},{\rm w}_0\}$. To do so, we develop Eq.~\eqref{eq:degeneracy} by computing the derivatives of $\xi_K$, which yields
\begin{equation}\label{eq:degeneracyrelation}
    \left[\frac{1}{H}\left(\frac{\dd H}{\dd z}+\frac{2}{\chi}\right)\right]_{\rm model1}= \left[\frac{1}{H}\left(\frac{\dd H}{\dd z}+\frac{2}{\chi}\right)\right]_{\rm model2}.
\end{equation}
For a full degeneracy of the parameters, this condition has to be verified for every redshift, and this is, of course, too strong of an assumption in our case. For that reason, the degeneracy is not exact, and the fiducial model is still favored. But it explains the shape of the contours, as shown in Fig.~\ref{fig:degeneracy}, where we plot three curves in parameter space along which the relation of Eq.~\eqref{eq:degeneracyrelation} is verified for three specified redshifts. We compare these lines to $\chi^2$ contours that reproduce the results of the MCMC. We see that the correlation corresponds to these degeneracy lines as long as they are close to each other. We can identify that the degeneracy at redshifts around $1$ is favored. Thus, it is the typical redshift to which we are sensitive with this approach.

\begin{figure}[!ht]
   \begin{center}
    \includegraphics[width=\columnwidth]{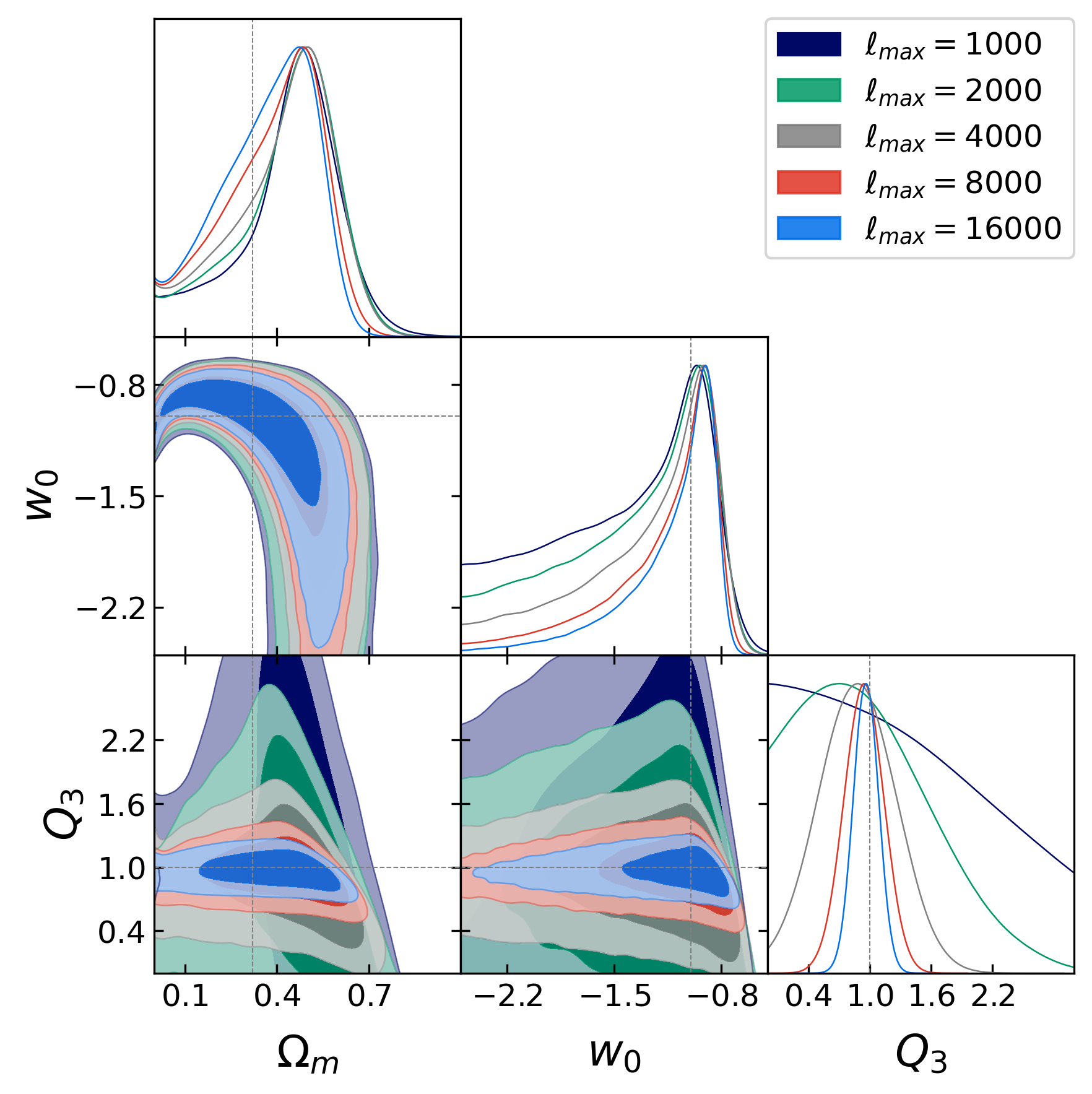}
  \end{center}
  \caption{Evolution of the triangle plot for sampled parameters $\{\Omega_{\rm m},{\rm w}_0,Q_3\}$ when the maximum used scale $\ell$ evolves from $1000$ to $16,000$. Dashed lines show the fiducial values of parameters.}
  \label{fig:MCMClmax}
\end{figure}

We then investigate the evolution of performance for different maximum scales $\ell_{max}=\{1000$, $2000$, $4000$, $8000$, $16,000\}$. We show the triangle plots comparison in Fig.~\ref{fig:MCMClmax}, and for completeness, the confidence intervals obtained for each case are detailed in Table \ref{tab:lmax}. We see, in agreement with Fig.~\ref{fig:signalnoisew0B}, that the knowledge of the bispectrum requires $\ell_{max} \geq 4000$. We then gain in precision when increasing $\ell_{max}$. On the other hand, we see that we gain little information on $\{\Omega_{\rm m},{\rm w}_0\}$ when increasing $\ell_{max}$ from $8000$ to $16,000$, most of the gain is on $Q_3$ in that case.

\begin{table}[!h]
\caption{\label{tab:lmax} Confidence interval evolution with $\ell_{max}$.}
\begin{ruledtabular}
\begin{tabular}{cccc}
$\ell_{max}$ & $\Omega_{\rm m}$ & ${\rm w}_0$ & $Q_3$ \\
\colrule
$1000$ & $0.440^{+0.19}_{-0.099}$ & $-1.62^{+0.85}_{-0.39}$ & $< 1.68$ \\
$2000$ & $0.433^{+0.20}_{-0.097}$ & $-1.52^{+0.77}_{-0.29}$ & $0.98^{+0.30}_{-0.94}$ \\
$4000$ & $0.42^{+0.20}_{-0.11}$ & $-1.38^{+0.63}_{-0.18}$ & $0.89\pm 0.37$ \\
$8000$ & $0.39^{+0.20}_{-0.11}$ & $-1.25^{+0.48}_{-0.10}$ & $0.94\pm 0.20$ \\
$16,000$ & $0.37^{+0.19}_{-0.11}$ & $-1.17^{+0.38}_{-0.075}$ & $0.96\pm 0.12$ \\
\end{tabular}
\end{ruledtabular}
\end{table}

\section{Photometric redshift errors}\label{sec:photometric}

In this section, we investigate both the effects of photometric redshift errors on cosmological constraints obtained from nulling and the ability of nulling in helping to control these systematics.

\subsection{Mean redshift errors}

The mean redshift errors correspond to the $10$ parameters $\zb{i}$ where $1 \leq i \leq 10$ that appear\footnote{Let us remind the reader that the method is insensitive to small variation of the redshift distribution of the low-redshift galaxy samples.} in Eq.~\eqref{eq:photoprob}. Until now, all of these were fixed to $0$. In this section, we add $10$ parameters to the system by considering the photometric mean redshift errors for the redshift distributions of the sources. The effect of having $\zb{i} \neq 0$ will mainly be a translation of the distribution of sources in bin $i$. This will affect the lensing kernel and thus the nulled lensing kernel by modifying its support. Such a modification could break nulling, and we should then be able to detect it. Thus, we investigate the effect of this systematic using Fisher formalism and MCMC tests. The nulling coefficients will still be computed for $\zb{}=0$, but we will consider the variations of $\mathcal{C}^{\kappa {\rm g}}_{ji}(\ell)$ with $\zb{}$ such that
\begin{equation}
    \frac{\dd \mathcal{C}^{\BNT{\kappa} {\rm g}}_{ai}(\ell)}{\dd \zb{k}}=p_{aj}(\boldsymbol{\theta})\frac{\dd \mathcal{C}^{\kappa {\rm g}}_{ji}(\ell)}{\dd \zb{k}}.
\end{equation}

However, one should be aware of the difference between $\sigma_z$ defined in Sec.~\ref{sssec:sources}, which gives the photometric redshift precision for one galaxy and the mean redshift shift or uncertainty $z_b$ on a redshift bin. Indeed, due to standard error reduction when averaging on large samples, the latter tends to be $1$ order of magnitude smaller than $\sigma_z$. This is the case in KiDS for which the mean redshift errors are known with precision better than $10^{-2}$ as shown in Ref.~\cite{Hildebrandt:2020rno}. Euclid's requirement is a $10^{-3}$ precision on mean redshift shifts \cite{EUCLID:2011zbd}.

\subsection{Discussion on the number of parameters}

The nulling matrix used as a probe in itself gives a number of constraints that are easy to determine. Indeed, for a tomographic construction with $N_z$ redshift bins, the BNT transform matrix is fully determined from $N_z-3$ independent coefficients. In our case, we have $10$ bins, so we find seven independent coefficients that are sufficient to determine the matrix. Thus, nulling gives $N_z-3$ independent observables such that, along with the possibility of cosmological degeneracy, one could constrain $N_z-3$ parameters.

The main systematic effects that we have to deal with are photometric redshift errors. We have $N_z$ of these, but the first one is ignored if we choose as an observable the cross-spectra between nulled shear and galaxy counts as described in Sec.~\ref{ssec:crossspectra}, allowing $N_z-1$ photometric redshift errors. As we cannot simultaneously constrain the cosmology and $N_z-1$ photometric redshift errors with only $N_z-3$ constraints, we should investigate the effect of photometric redshift errors as systematics. However, conversely, if we suppose that we have precise enough constraints on cosmological parameters given by a survey, we could bring $N_z-3$ constraints to the $N_z-1$ photometric redshift errors.

\subsection{Joint degeneracies between bin mean redshifts and cosmological parameters}

As stated before, with a 10 tomographic bins setting, only seven constraints can be obtained with the nulling technique. If we further add three cosmological parameters such as $\{\Omega_{\rm m},{\rm w}_0,{\rm w}_a\}$, we are left with an ensemble of 10 variables in which five combinations cannot be constrained. Note that $Q_3$ is not considered here as its value does not influence the computation of the nulling coefficients. We can search for such degeneracies by identifying the combination of parameters giving the same nulling coefficients $p_{ai}$ to first order in parameters variations. It can be expressed as
\begin{equation}
    \delta p_{ai} = \sum_\theta \frac{\partial p_{ai}}{\partial \theta} \delta \theta = 0,
\end{equation}
where $p_{ai}$ is a nulling coefficient with $a>2$ and $i=a-2$ or $i=a-1$ and the sum is made over all parameters. Note that for each $a$, given the constraint on each line, $p_{a,a-2}+p_{a,a-1}+p_{a,a}=0$, only one 
parameter can vary independently that we choose to be $p_a=p_{a,a-2}$.

Ignoring $\zb{1}$ which here is a free parameter and restricting $\theta$ to cosmological parameters $\{\Omega_{\rm m},{\rm w}_0,{\rm w}_a\}$ (one could add $\Omega_K$ in a nonflat universe), we find a double recurrence relation defining possible mean redshift errors that would cancel a deviation of cosmological parameters from the fiducial model in terms of nulling coefficient as 
\begin{align}\label{variationpa}
    \delta p_{a}=&\frac{\partial p_{a}}{\partial \zb{a-2}} \delta \zb{a-2}+\frac{\partial p_{a}}{\partial \zb{a-1}} \delta \zb{a-1}+\frac{\partial p_{a}}{\partial \zb{a}} \delta \zb{a}\nonumber \\
    & +\sum_{\theta_c}  \frac{\partial p_{a}}{\partial \theta_c} \delta \theta_c
\end{align}
where $a \geq 4$ with $\theta_c \in \{\Omega_{\rm m},{\rm w}_0,{\rm w}_a\}$. The consequences of this relation are multifold. When the cosmological parameters are fixed, $\delta \theta_c=0$, there is a relation between $\delta \zb{a-2}$, $\delta \zb{a-1}$ and $\delta \zb{a}$ that makes $\delta p_{a}$ vanish. Then, for any arbitrary variations of $\delta\zb{2}$ and $\delta\zb{3}$, it is possible to find a set of values for $\zb{a}$ that make the entire transformation matrix invariant. Basically, nulling can be preserved if arbitrary errors for two bins are compensated by errors in the others. Together with the freedom in the mean redshift of the first bin, we are left with a kernel of dimension $3$ for the mean redshift alone.

From Eq.~\eqref{variationpa}, we also determine that the null test for $N_z$ redshift bins brings $N_z-3$ constraints, so seven in our case, which implies some possible degeneracies. This can be indeed observed in the $12 \times 12$ Fisher Matrix for the set of parameters $\{\Omega_{\rm m},{\rm w}_0,{\rm w}_a\}$ and $\{\zb{i} \rm{\ for\ } 2\leq i \leq 10\}$ by following Eq.~\eqref{eq:FoM}. While still working in a flat Universe, we choose to be more complete on dark energy and bring ${\rm w}_a$ back as a parameter but ignore $Q_3$ as it is to be considered as a nuisance parameter.\footnote{One could still consider it and marginalize over it, and results should be similar, but this would require two matrix inversions adding possible numerical instability.} The diagonalization of the resulting  Fisher Matrix clearly exhibits two subeigenspaces of respective dimension $7$ and $5$ with eigenvalues of order $10^5$—$10^6$ for the first dimension space and $10^{-9}$—$10^2$ for the second. 

Note that we find similarly a seven-dimension space with eigenvalues of order $10^5$—$10^6$ and a seven-dimension space with eigenvalues of order $0$—$10^2$, when working with the full $14 \times 14$ Fisher Matrix for 10 mean redshifts and the set of parameters $\{\Omega_{\rm m},{\rm w}_0,{\rm w}_a,\Omega_K,Q_3\}$.
 
\subsection{Exploring the impact of priors on the bin mean redshifts}

As we cannot simultaneously constrain cosmology and photometric redshift errors, it is useful to quantify the precision requirement on photometric redshifts to preserve the obtained constraint on cosmological parameters. 
A Fisher analysis shows that, to avoid less than $10 \%$ loss in value of the FOM, a precision on $\zb{}$ in the $4\times 10^{-4}$ range is required. We analyze here more precisely the impact of a degradation of precision on these parameters.

\begin{figure}[!ht]
   \begin{center}
    \includegraphics[width=\columnwidth]{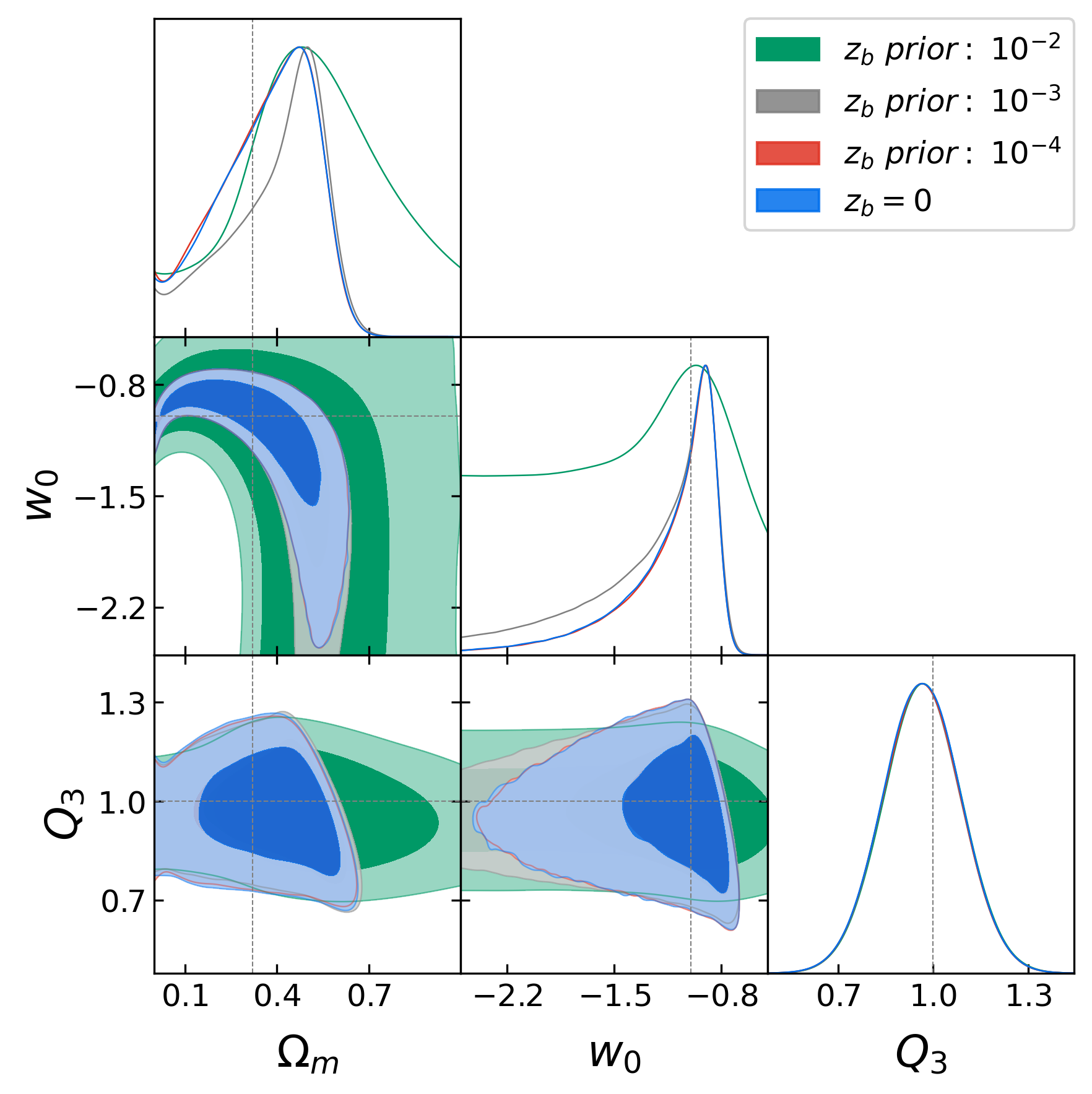}
  \end{center}
  \caption{Evolution of the triangle plot for sampled parameters $\{\Omega_{\rm m},{\rm w}_0,Q_3\}$ for Gaussian prior scales in $\{10^{-2}$, $10^{-3},10^{-4},0\}$. Dashed lines show the fiducial values of parameters.}
  \label{fig:MCMCzb}
\end{figure}

To do so, we introduce priors on $\zb{i}$ and investigate the effect of adding these 10 parameters along with their priors to the MCMC setting. For the prior, we choose 10 independent Gaussian distributions, one for each mean redshift error, centered on $0$, and we vary the scale $\sigma_{\zb{}}$, which would be the photometric mean redshift accuracy. We choose to test for $\sigma_{\zb{}}=\{10^{-4},10^{-3},10^{-2}\}$ as these values cover the current accuracy of Stage III surveys with KiDS having $z_b$ of order $10^{-2}$ \cite{Hildebrandt:2020rno}, the aim of Stage IV surveys as Euclid's requirement is set at $z_b < 10^{-3}$ \cite{EUCLID:2011zbd}. Then, $z_b$ of order $10^{-4}$ could be an optimistic estimate of future precision or a gain of mean redshift precision obtained using the BNT transform as a probe. As $\sigma_{\zb{}}$ is small, we only consider the two first orders in $\zb{}$ for the correction to $\mathcal{C}^{\kappa {\rm g}}_{ji}(\ell)$ in the actual computation. This preserves the computation time of the MCMC run by preventing a new computation of $\mathcal{C}^{\kappa {\rm g}}_{ji}(\ell)$ at each step. We show in Fig.~\ref{fig:MCMCzb} the evolution of the triangle plot with the prior scale where $\sigma_{\zb{}}=0$ stands for perfectly known mean redshifts and corresponds to Fig.~\ref{fig:MCMC}.

We can conclude that the constraints are preserved for $\sigma_{\zb{}} <  10^{-3}$, which means that the source redshifts should be known up to $10^{-3}$ to fully exploit the nulling correlation between $\Omega_{\rm m}$ and ${\rm w}_0$. We then show the actual sampled confidence intervals for each prior in Table~\ref{tab:zb}.

\begin{table}[!h]
\caption{\label{tab:zb} Evolution of confidence intervals with priors on $\zb{}$.}
\begin{ruledtabular}
\begin{tabular}{ccccc}
$\zb{}$ prior & $\Omega_{\rm m}$ & ${\rm w}_0$ & $Q_3$ \\
\colrule
$10^{-2}$ & $0.52\pm 0.22$ & $<-1.07$ & $0.97\pm 0.12$ \\
$10^{-3}$ & $0.396^{+0.19}_{-0.092}$ & $-1.32^{+0.55}_{-0.13}$ & $0.96\pm 0.12$ \\
$10^{-4}$ & $0.36^{+0.19}_{-0.11}$ & $-1.17^{+0.38}_{-0.073}$ & $0.96\pm 0.12$ \\
$0$      & $0.37^{+0.19}_{-0.11}$ & $-1.17^{+0.38}_{-0.075}$ & $0.96\pm 0.12$ \\
\end{tabular}
\end{ruledtabular}
\end{table}

Another valid conclusion on photometric redshift systematics would be to consider the case of probe combination. In this context, the cosmological parameters and the cross-spectra could be considered to be determined via a so-called 3$\times$2pt analysis, and the nulling observable could then be seen as additional constraints on photometric mean redshift errors. Thus, it would bring a valuable gain when dealing with systematics.

\section{Conclusion}

We present here a proof of concept of a novel probe that can offer complementary constraints on cosmological parameters from tomographic cosmic shear observations. It is based on nulling properties exploiting the BNT transform's sensitivity only to the geometry of the background space-time. We detail the construction of such a new observable, its general methodology, and how it can be used to exploit the statistical power of cosmic shear maps even at very small physical scales while being entirely independent of physical assumptions on how density fluctuations develop at small scales. The constraints this method can bring are then totally independent of assumptions regarding the nonlinear growth of structures, the impact of baryon physics, galaxy biasing and its evolution with redshift, and also the presence of intrinsic alignments. Departure from nulling is only expected to arise due to subleading effects to the weak lensing regime (i.e., transition to strong lensing) that can affect the very small angular scales. In this regard, the introduction of bispectrum corrections that describe the first correction to the weak lensing regime allows us to safely identify the scales that can be exploited for nulling.

We examine the performance of this method through comprehensive analysis using Fisher Matrix forecasts and MCMC sampling. In particular, it allows us to assess the impact of various systematics and corrections such as the reduced shear, the magnification bias, and photometric redshift precision.

Our results illustrate the potential of nulling in enhancing the precision of cosmological measurements, particularly  the total matter density $\Omega_{\rm m}$ and dark energy equation of state, to which the BNT transform coefficients are mostly sensitive. This work reveals notably that, in Euclid's context, the combination of nulling with the 3$\times$2pt method could bring new constraints on the dark energy equation of state as we exploit properties of weak lensing that are currently unused. Moreover, we identified a precision requirement for photometric redshifts of the order of $10^{-3}$ in our current setting, needed to reach such results. However, considering the forecasts for the Euclid survey, we conclude that cosmic shear nulling may not significantly enhance parameter constraints beyond those obtained from the main probes of Stage IV surveys but rather be used as a consistency check. In addition, we explicitly demonstrate that this probe is highly sensitive to photometric redshift uncertainties. That is why we conclude that, in the short term, nulling would be most efficient as an additional probe or consistency check for controlling systematics of photometric redshift calibration. Indeed, nulling provides us with a particularly clean and robust diagnostic of photometric redshift systematics, especially at high redshift, where external calibration is most challenging. In this sense, it would be an indirect improvement of precision on cosmological parameters through the reduction of systematic effects and uncertainties on photometric redshifts.

Based on these results, this methodology is clearly a promising addition to the current analysis toolkit of weak lensing and galaxy surveys. Through its computational simplicity and its dependence only on a reduced number of cosmological parameters, it is a virtually costfree and robust probe that introduces no additional systematics and can help reduce existing systematic effects and uncertainties. In addition to the  lensing shear ratio approach, which has been exploited in DES, cosmic shear nulling could be used in future surveys as an additional probe or as a consistency test. 

Let us finally note that the methodology used in this paper has the advantage of being simple but does not guarantee that it fully extracts the information contained in the nulling condition. Indeed, we have chosen one specific tracer (in our case, galaxy distributions) and one specific estimator (in our case, the cross-correlation between the low-redshift tracers and the nulled lensing maps). Future works could naturally extend this nulling test to other tracers. One other avenue of research could be to try and go beyond the condition on the cross-correlation which only tests the linear decorrelation between the low-redshift tracers and the nulled lensing maps. However, testing full statistical independence between the maps is notoriously difficult, particularly when it comes to accurately modeling the noise. We leave this for future work.

\section{Data Availability}

The data that support the findings of this article are openly available \cite{touzeau_2026_19697500}.

\section{Acknowledgements}This work has received funding from the Centre National d’Etudes Spatiales for travel and hardware. It was carried out using the Infinity Cluster hosted by the Institut d’Astrophysique de Paris. We also acknowledge the Euclid consortium for organizing events and for providing opportunities to present this work at various stages, and we thank several of its members for useful discussions. S.C. warmly thanks Helene Dupuy-Velu for her contributions to an earlier version of this work.

\appendix
\section{Conventions and useful equations}
\subsection{Fourier transform}

In this paper, the following conventions are used for $n$-dimensional Fourier transforms
\begin{align}
\tilde{f}(\boldsymbol{k}) &= \frac{1}{\sqrt{(2 \pi)^n}} \int \dd^nx e^{-i \boldsymbol{k}\cdot\boldsymbol{x}} f(\boldsymbol{x}), \\
f(\boldsymbol{x}) &= \frac{1}{\sqrt{(2 \pi)^n}} \int \dd^nk e^{i \boldsymbol{k}\cdot\boldsymbol{x}} \tilde{f}(\boldsymbol{k}).
\end{align}
For simplicity, we will drop the tilde in the main text. The variables we will use will be $\mathbf{n}$ for the two-dimensional unit less position in the sky and $\boldsymbol{\ell}$ for the corresponding scale. For the usual three-dimension space, we will use $\chi$ [or $D_K(\chi)$ in curved space] for distances, and the wave vector is $\boldsymbol{k}$.

\subsection{Polyspectra normalizations}

We use the following conventions for spectrum and bispectrum:

\subsubsection{Spectrum and two-point correlation functions}
\begin{align}
\left<\tilde{\delta}_{\rm m}(\boldsymbol{k})\tilde{\delta}_{\rm m}(\boldsymbol{k}')\right>&=\delta_D(\boldsymbol{k}+\boldsymbol{k}')P(k) \\
\left<\delta_{\rm m}(\boldsymbol{x})\delta_{\rm m}(\boldsymbol{x}+\boldsymbol{r})\right>&= \int \frac{d^3 k}{(2 \pi)^3} P(k)e^{i \boldsymbol{k} . \boldsymbol{r}} \\
\sigma^2 = \left<\delta_{\rm m}^2\right> &= \xi(0) = \int \frac{d^3k}{(2 \pi)^3} P(k) \\
\frac{d\sigma^2}{d{\rm ln}k}&=\frac{4 \pi k^3}{(2 \pi)^3} P(k)
\end{align}

\subsubsection{Bispectrum}
{\footnotesize\begin{equation}
\left<\tilde{\delta}_{\rm m}(\boldsymbol{k}_1)\tilde{\delta}_{\rm m}(\boldsymbol{k}_2)\tilde{\delta}_{\rm m}(\boldsymbol{k}_3)\right> = \frac{1}{\sqrt{(2 \pi)^3}} \delta_D(\boldsymbol{k}_1 + \boldsymbol{k}_2 + \boldsymbol{k}_3) B(\boldsymbol{k}_1,\boldsymbol{k}_2,\boldsymbol{k}_3).
\end{equation}}
And using Eq.~\eqref{eq:bispectrum}, we compute
\begin{align}
\left<\delta_{\rm m}^3\right> &= 3 Q_3 \int \frac{d^3k_1}{(2 \pi)^3} \frac{d^3k_2}{(2 \pi)^3} P(k_1) P(k_2) \\
&= 3 Q_3  \left<\delta_{\rm m}^2\right>^2
\end{align}
such that $S_3=3Q_3$.

\section{Proof of cosmological degeneracy}\label{sec:proofdegen}

Let us give here the proof of the cosmological degeneracy in the nulling coefficients.
It can first be noticed that the nulling procedure can be built from a set of arbitrarily small bins since we can always build the $N_z/2$ case coefficients from the $N_z$ case coefficients.\footnote{More precisely, the values of the $p^{(n_z/2)}_{ij}$ elements are obtained from the $p^{(n_z)}_{ij}$ through a transformation matrix $T$ by imposing that $T_{ij}\,p^{(nz)}_{j2i}=T_{ij}\,p^{(nz)}_{j2i-1}=1$, $T_{ij}\,p^{(nz)}_{j2i-2}=T_{ij}\,p^{(nz)}_{j2i-3}$, $T_{ij}\,p^{(n_{z})}_{j2i-4}=T_{ij}\,p^{(n_{z})}_{j2i-5}$, and $T_{j,2(j-2)}=T_{j,2(j-2)-1}=0$, which is always possible (six equations for eight coefficients). Then, we have $p^{(n_{z}/2)}_{ik}=T_{ij}\,p^{(n_{z})}_{j2k}=T_{ij}\,p^{(n_{z})}_{j2k-1}$.}

We then consider infinitely small bins. We define the bin width as $\Delta\chi(\chi)$ and redefine the moments from Eq.~\eqref{eq:moments} as
\begin{equation}
    n^{(0)}(\chi)=\int_\chi^{\chi +\Delta\chi(\chi)} \dd\chi' n(\chi')=1,
\end{equation}
which allows us to implicitly define $\Delta\chi(\chi)$ and
\begin{equation}
n^{(1)}(\chi)=\int_\chi^{\chi +\Delta\chi(\chi)} \dd\chi' n(\chi')\xi_K(\chi'),
\end{equation}
where $\xi_K=1/F_K$. We then derive the following relations by derivation with respect to $\chi$,
\begin{equation}
    n(\chi)=n(\chi+\Delta\chi(\chi))\left[1+\frac{\dd}{\dd\chi}\Delta\chi(\chi) \right],
\end{equation}
such that
\begin{align}
    \frac{\dd n^{(1)}(\chi)}{\dd \chi} &= n(\chi) [\xi_K(\chi+\Delta\chi)-\xi_K(\chi)] \\
    &\approx n(\chi)\xi_K'(\chi)\Delta\chi,
\end{align}
and (noticing that $n'(\chi)\Delta\chi(\chi)+n(\chi)\Delta\chi'(\chi)=0$),
\begin{equation}
    \frac{\dd^2 n^{(1)}(\chi)}{\dd \chi^2}=n(\chi)\xi_K''(\chi)\Delta\chi.
\end{equation}
We have
\begin{equation}
    p_{i-2,i}=\frac{n^{(1)\prime}(\chi_2)\Delta\chi_2 + n^{(1)\prime\prime}(\chi_2)(\Delta\chi_2)^2/2}{n^{(1)\prime}(\chi_1)\Delta\chi_1 + n^{(1)\prime\prime}(\chi_1)(\Delta\chi_1)^2/2}.
\end{equation}

We then develop at first order in $\Delta\chi_1$ noting that $\chi_2=\chi_1+\Delta\chi_1$ and $\Delta\chi_2=\Delta\chi_1+\Delta\chi'(\chi_1)\Delta\chi_1$ and replace $\chi=\chi_1$ and $\Delta\chi_1=\Delta\chi$. We obtain
\begin{equation}
    p_{i-2,i}=1+\frac{n^{(1)\prime\prime}(\chi)}{n^{(1)\prime}(\chi)}\Delta\chi-\frac{n'(\chi)}{n(\chi)}\Delta\chi
\end{equation}
which leads to
\begin{equation}
    p_{i-2,i}=1+\frac{\xi_K''(\chi)}{\xi_K'(\chi)}\Delta\chi-\frac{n'(\chi)}{n(\chi)}\Delta\chi.
\end{equation}

Since $n(z)\dd z=n(\chi)\dd\chi$, $\xi_K(z)=\xi_K(\chi(z))$ and $\Delta\chi=\frac{\dd\chi}{\dd z} \dd z$, we can change variable from $\chi$ to $z$ and find
\begin{equation}
    p_{i-2,i}=1+\frac{\dd^2\xi_K/\dd z^2}{\dd\xi_K/\dd z}\Delta z-\frac{1}{n(z)} \frac{\dd n(z)}{\dd z}\Delta z.
\end{equation}

One can easily identify two sources of invariance of the nulling coefficient with respect to parametrization. Indeed, the first term ($\xi_K''/\xi_K'$) traduces the dependence with cosmological parameters. The other is the dependence with the redshift distribution of galaxies (which is not a studied parameter, but a given data point in this work). We then see that two cosmological models will give the same nulling coefficients if we can find two constants $A$ and $B$ such that
\begin{equation}
    \xi_K^{\rm model1}=A\xi_K^{\rm model2}+B,
\end{equation}
which can be rewritten as
\begin{equation}
    \frac{1}{F_K^{\rm model1}}=\frac{A}{F_K^{\rm model2}}+B.
\end{equation}

\section{Low-redshift galaxy binning scheme} \label{sec:ggal}

As stated in Sec.~\ref{sssec:lowzg}, neither the galaxy distribution $n_i(z)$ nor the nulled lensing kernel $\BNT{w}_a(z)$ exactly reaches zero at low redshifts when $\sigma_z \neq 0$. Consequently, the overlap between $n_i(z)$ and $\BNT{w}_a(z)$ for $i \leq a-3$ must be treated as "noise." To minimize this noise, we impose a threshold that ensures that the two functions overlap only when their contributions are negligible. Specifically, we require that this overlap never occurs if one is greater than $10 \%$ of its maximum value.

This Appendix describes the binning scheme for low-redshift galaxy distributions, optimized to satisfy this condition. The approach aims to balance signal suppression and data retention, ensuring that unwanted correlations are minimized without overly reducing the available data.

\subsection{Methodology}

To ensure that the nulled lensing kernel $\BNT{w}_a(z)$ and the galaxy distribution $n_{a-3}(z)$ meet the $10\%$ threshold, we define two key redshifts:

1. The redshift $z_{\BNT{w}}(a)$, defined as the redshift below which the nulled lensing kernel $\BNT{w}_a(z)$ remains less than $10\%$ of its maximum value. Specifically, we require
\[
\BNT{w}_a(z) \leq 0.1 \times \max(\BNT{w}_a(z)) \approx 0.02 \quad \text{for } z \leq z_{\BNT{w}}(a),
\]
where we take $0.2$ as an approximated common maximum value across all nulled lensing kernels. Additionally, $z_{\BNT{w}}(a)$ is chosen on the left of the curve, avoiding the higher-redshift region where the kernel drops below $10\%$ again.

2. The photometric redshift $z_{\rm g}(a-3)$, defined as the redshift where the galaxy bin is cut to ensure that the galaxy number density $n_{a-3}$ is sufficiently small at $z_{\BNT{w}}(a)$ and higher redshifts. Specifically, we require
\[
n_{a-3}(z_{\BNT{w}}(a)) \leq 0.1 \times n(z),
\]
to ensure that the overlap between $n_{a-3}(z)$ and $\BNT{w}_a(z)$ contributes negligibly to the signal.

\subsection{Implementation details}

The bin limits $z_{\rm g}(i)$ are chosen to meet the conditions
\begin{align*}
z_{\rm g}(1) &= z_{\rm min}, \\
z_{\rm g}(i \geq 8) &= z_i, \\
f_i(z_{\BNT{w}}(i+3)) &= 0.1 \quad \text{for } i \leq 7,
\end{align*}
where $f_i(z)=\int_{z_{\rm g}(i)}^{z_{\rm g}(i+1)} \dd z_p\, p_{\rm ph}(z_p,z,\sigma_z,\zb{i}=0)$ is the fraction of galaxies at real redshift $z$ that are labeled in bin $i$, due to the measured photometric redshift; see Eq.~\eqref{eq:lowzn(z)} for details. With these bin limits $\{z_{\rm g}(i)\}$, the galaxy distribution $n^{\rm g}_i(z)$ is constructed and the overlap condition is satisfied such that for $a \geq 4$, $n^{\rm g}_{a-3}(z)$ and $\BNT{w}_a(z)$ only overlap when both are below $10\%$ of their respective maximum values. This ensures that their product contributes less than $1\%$ of the usual non-nulled signal.

\subsection{Results}

The resulting bin limits for the low-redshift galaxy distributions are summarized in Table~\ref{tab:bins}. The binning is optimized for different values of the photometric redshift uncertainty $\sigma_z$.

\begin{table}[!h]
\caption{\label{tab:bins} Low-redshift galaxy bin limits.}
\begin{ruledtabular}
\begin{tabular}{ccccc}
 Source & & $\sigma_z = 0$ & $\sigma_z = 0.02$ & $\sigma_z = 0.05$ \\
 \colrule
 $0.001$ & $z_{\rm g}(1)$ & $0.001$ & $0.001$ & $0.001$ \\
 $0.417733$ & $z_{\rm g}(2)$ & $0.461323$ & $0.415625$ & $0.329181$ \\
 $0.559519$ & $z_{\rm g}(3)$ & $0.594594$ & $0.540256$ & $0.438217$ \\
 $0.677054$ & $z_{\rm g}(4)$ & $0.708882$ & $0.647685$ & $0.534286$ \\
 $0.787459$ & $z_{\rm g}(5)$ & $0.817948$ & $0.750639$ & $0.62635$ \\
 $0.898769$ & $z_{\rm g}(6)$ & $0.928984$ & $0.855755$ & $0.71946$ \\
 $1.01771$ & $z_{\rm g}(7)$ & $1.04815$ & $0.968783$ & $0.818249$ \\
 $1.15317$ & $z_{\rm g}(8)$ & $1.18283$ & $1.09551$ & $0.92644$ \\
 $1.32208$ & $z_{\rm g}(9)$ & $1.32208$ & $1.32208$ & $1.32208$ \\
 $1.574$ & $z_{\rm g}(10)$ & $1.574$ & $1.574$ & $1.574$ \\
 $2.5$ & $z_{\rm g}(11)$ & $2.5$ & $2.5$ & $2.5$ \\ 
\end{tabular}
\end{ruledtabular}
\end{table}

\section{Magnification bias slope fitting formula and values}\label{sec:MBSlope}

We give here the fitting formula for the magnification bias slope as
\begin{equation}
    b_M(z)=b_{M0}+b_{M1}(z)+b_{M2}z^2+b_{M3}z^3,
\end{equation}
where $b_M(z)=5s(z)-2$ and 
\begin{align}
    b_{M0}&=-1.50685, \\
    b_{M1}&=1.35034, \\
    b_{M2}&=0.08321, \\
    b_{M3}&=0.04279.
\end{align}

These values are obtained by fitting on the fiducial magnification bias of the Euclid Flagship simulation described in Ref.~\cite{Euclid:2024few}.

Then, we determine $s_i$ in source bin $i$ or $s_i^{\rm g}$ in low-redshift bin $i$ by evaluating $s(z)$ at the mean redshift of the galaxy distribution bin
\begin{align}
    s_i&=s\left(\frac{\int \dd z z n_i(z)}{\int \dd z n_i(z)}\right) \\
    s_i^{\rm g}&=s\left(\frac{\int \dd z z n_i^{\rm g}(z)}{\int \dd z n_i^{\rm g}(z)}\right).
\end{align}
The values of the slope are given in Tables~\ref{tab:sslope} and~\ref{tab:gslope}.

\begin{table}[!ht]
\caption{\label{tab:sslope} Slope $s$ for source redshift bins.}
\begin{ruledtabular}
\begin{tabular}{cccc}
$i$ & $s_i(\sigma_z=0)$& $s_i(\sigma_z=0.02)$& $s_i(\sigma_z=0.05)$\\
\colrule
$1$ & $0.227984$ & $0.187187$ & $0.189713$ \\
$2$ & $0.244983$ & $0.237492$ & $0.240782$ \\
$3$ & $0.275786$ & $0.27484$ & $0.277194$ \\
$4$ & $0.307159$ & $0.309031$ & $0.310439$ \\
$5$ & $0.339176$ & $0.343247$ & $0.343648$ \\
$6$ & $0.373195$ & $0.379737$ & $0.378987$ \\
$7$ & $0.411148$ & $0.421163$ & $0.418999$ \\
$8$ & $0.456225$ & $0.472104$ & $0.468077$ \\
$9$ & $0.515165$ & $0.543689$ & $0.537139$ \\
$10$ & $0.608531$ & $0.681767$ & $0.675039$ \\
\end{tabular}
\end{ruledtabular}
\end{table}

\begin{table}[!ht]
\caption{\label{tab:gslope} Slope $s$ for low-redshift galaxy bins.}
\begin{ruledtabular}
\begin{tabular}{cccc}
$i$ & $s_i^{\rm g}(\sigma_z=0)$& $s_i^{\rm g}(\sigma_z=0.02)$& $s_i^{\rm g}(\sigma_z=0.05)$\\
\colrule
$1$ & $0.235984$ & $0.18676$ & $0.172655$ \\
$2$ & $0.253473$ & $0.234131$ & $0.211573$ \\
$3$ & $0.284748$ & $0.267585$ & $0.239583$ \\
$4$ & $0.316065$ & $0.298972$ & $0.266159$ \\
$5$ & $0.348152$ & $0.330827$ & $0.292951$ \\
$6$ & $0.382406$ & $0.36506$ & $0.321312$ \\
$7$ & $0.420624$ & $0.403675$ & $0.352552$ \\
$8$ & $0.464239$ & $0.460772$ & $0.425731$ \\
$9$ & $0.515165$ & $0.543689$ & $0.537139$ \\
$10$ & $0.608531$ & $0.681767$ & $0.675039$ \\
\end{tabular}
\end{ruledtabular}
\end{table}

\section{Lensing shear ratio computations on observables}\label{sec:lsr}

We investigate here the so-called lensing shear ratio observable, as defined in Refs.~\cite{DES:2018lpj,DES:2021jzg}, to further compare it with the BNT nulling approach presented here. The lensing shear ratio observable is based on the observation that, when the distance distribution of the low-redshift galaxies is narrow enough, the cross-spectra  ${\mathcal{C}^{\kappa {\rm g}}_{ji}(\ell)}/{\mathcal{C}^{\kappa {\rm g}}_{ki}(\ell)}$ are expected to be scale independent irrespectively of behavior of the matter power spectra. Under the assumption that the distances of the low-redshift tracers can be replaced by their average in the computation of the cross-spectra, the lensing shear ratio defined in Eq.~\eqref{eq:lsr} is written as
\begin{equation}
    q^{i,j,k}=\frac{\int \dd z_j n_j(z_j) w(\chi(z_j),\chi(z_i))}{\int \dd z_k n_k(z_k) w(\chi(z_k),\chi(z_i))}
\end{equation}
where $z_i$ is the mean redshift of the low-redshift galaxy in bin $i$ and $n_j$ and $n_k$ are the source redshift distributions in source bins $j$ and $k$. This quantity can then be further written,
\begin{equation}
    q^{i,j,k}=\frac{\int \dd \chi_j n_j(\chi_j) \left(1/F_K(\chi_j) - 1/F_K(\chi_i)\right)}{\int \dd \chi_k n_k(\chi_k) \left(1/F_K(\chi_k) - 1/F_K(\chi_i)\right)},
\end{equation}
where $\chi_i=\chi(z_i)$. We can then express these moments in terms of the moments defined in Eq.~\eqref{eq:moments},
\begin{equation}
    q^{i,j,k}=\frac{n_j^{(1)} - n_j^{(0)}/F_K(\chi_i)}{n_k^{(1)} - n_k^{(0)}/F_K(\chi_i)},
\end{equation}
which simplifies in the case of equipopulated source bins as
\begin{equation}
    q^{i,j,k}=\frac{n_j^{(1)} - 1/F_K(\chi_i)}{n_k^{(1)} - 1/F_K(\chi_i)}.
\end{equation}

We recover here the same combination as for the nulling method based on BNT transforms. That implies in particular that the two methods are expected to be sensitive to the same degeneracies among cosmological models. In practice, however, constraints will come from slightly different redshift ranges, as the shear ratio method involves the redshift range of the low-redshift galaxies, whereas the BNT-based nulling approach provides constraints in the redshift range of the lenses only and is insensitive to the precise distances of the low-redshift tracers.

One can also see that the narrowness of the low-redshift galaxy bin $i$ and the precise knowledge of its mean redshift $z_i$ are key to the shear ratio method, whereas it is not for the BNT-based nulling method.

\section{Explanation of data}

Nontrivial or nonexplicit data used in this work are available in Ref.~\cite{touzeau_2026_19697500}.

For the MCMC sampling, data include chains, likelihoods and required synthetic data (covariance, data-vector: cross-spectra, corrections).

For the Fisher analysis, data include values for Figs.~\ref{fig:signalnoisew0} and~\ref{fig:signalnoisew0B} and $\chi 2$ tabulated values along with degeneracy lines coordinates for Fig.~\ref{fig:degeneracy}.

\bibliography{bibliography.bib}
\end{document}